\documentclass[twocolumn]{article}
\usepackage[margin=1in]{geometry}

\usepackage{graphicx}
\usepackage{hyperref} 
\usepackage{color}
\usepackage{amsmath} 
\usepackage{float}
\usepackage[super, numbers]{natbib} 
\usepackage[table]{xcolor} 
\usepackage{subcaption} 
\usepackage{booktabs} 
\usepackage{makecell} 
\usepackage{multirow} 
\usepackage{authblk}
\usepackage{setspace}

\usepackage{changepage} 
\newlength{\offsetpage}
\newenvironment{widepage}{\begin{adjustwidth}{-\offsetpage}{-\offsetpage}%
		\addtolength{\textwidth}{2\offsetpage}}%
	{\end{adjustwidth}} 

\hypersetup{ colorlinks,
	citecolor=blue,
	filecolor=blue,
	linkcolor=blue,
	urlcolor=blue
}

\begin{document}

\date{}

\title{Deep Learning for fully automatic detection, segmentation, and Gleason Grade estimation of prostate cancer in multiparametric Magnetic Resonance Images}

\author[1]{Oscar J. Pellicer-Valero}
\author[2]{José L. Marenco Jiménez}
\author[3]{Victor Gonzalez-Perez}
\author[2]{Juan Luis Casanova Ramón-Borja}

\author[4]{Isabel Martín García}
\author[4]{María Barrios Benito}
\author[4]{Paula Pelechano Gómez}

\author[2]{José Rubio-Briones}
\author[5]{María José Rupérez}
\author[1]{José D. Martín-Guerrero}

\affil[1]{Intelligent Data Analysis Laboratory, Department of Electronic Engineering, ETSE (Engineering School), Universitat de València (UV), Av. Universitat, sn, 46100 Bujassot, Valencia, Spain. E-Mails: \texttt{Oscar.Pellicer@uv.es (+34 9635 44022)}, \texttt{jose.d.martin@uv.es}}

\affil[2]{Department of Urology, Fundación Instituto Valenciano de Oncología (FIVO), Beltrán Báguena, 8, 46009 Valencia, Spain. E-Mails: \texttt{jlmarencoj@gmail.com}, \texttt{jcasanova@fivo.org}, \texttt{jrubio@fivo.org}}

\affil[3]{Department of Medical Physics, Fundación Instituto Valenciano de Oncología (FIVO), Beltrán Báguena, 8, 46009 Valencia, Spain. E-Mails: \texttt{vgonzalezper@hotmail.com}}

\affil[4]{Department of Radiodiagnosis, Fundación Instituto Valenciano de Oncología (FIVO), Beltrán Báguena, 8, 46009 Valencia, Spain. E-Mails: \texttt{mismaga99@gmail.com}, \texttt{mar7esc@gmail.com}, \texttt{ppelechano@hotmail.com}}

\affil[5]{Instituto de Ingeniería Mecánica y Biomecánica, Universitat Politècnica de València (UPV), Camino de Vera, sn, 46022, Valencia, Spain. E-Mail: \texttt{mjrupere@upvnet.upv.es}}

\maketitle

\begin{abstract}\label{abstract}

The emergence of multi-parametric magnetic resonance imaging (mpMRI) has had a profound impact on the diagnosis of prostate cancers (PCa), which is the most prevalent malignancy in males in the western world, enabling a better selection of patients for confirmation biopsy. However, analyzing these images is complex even for experts, hence opening an opportunity for computer-aided diagnosis systems to seize. This paper proposes a fully automatic system based on Deep Learning that takes a prostate mpMRI from a PCa-suspect patient and, by leveraging the Retina U-Net detection framework, locates PCa lesions, segments them, and predicts their most likely Gleason grade group (GGG). It uses 490 mpMRIs for training/validation, and 75 patients for testing from two different datasets: ProstateX and IVO (Valencia Oncology Institute Foundation). In the test set, it achieves an excellent lesion-level AUC/sensitivity/specificity for the GGG$\geq$2 significance criterion of 0.96/1.00/0.79 for the ProstateX dataset, and 0.95/1.00/0.80 for the IVO dataset. Evaluated at a patient level, the results are 0.87/1.00/0.375 in ProstateX, and 0.91/1.00/0.762 in IVO. Furthermore, on the online ProstateX grand challenge, the model obtained an AUC of 0.85 (0.87 when trained only on the ProstateX data, tying up with the original winner of the challenge). For expert comparison, IVO radiologist's PI-RADS 4 sensitivity/specificity were 0.88/0.56 at a lesion level, and 0.85/0.58 at a patient level. Additional subsystems for automatic prostate zonal segmentation and mpMRI non-rigid sequence registration were also employed to produce the final fully automated system. The code for the ProstateX-trained system has been made openly available at \url{https://github.com/OscarPellicer/prostate_lesion_detection}. We hope that this will represent a landmark for future research to use, compare and improve upon.

	
\end{abstract}

\section{Introduction}\label{introduction}

Prostate cancer (PCa) is the most frequently diagnosed malignancy in males in Europe and the USA and the second in the number of deaths\cite{Bray2018}. The introduction of multiparametric magnetic resonance imaging (mpMRI) has drastically changed the diagnostic approach of PCa: whereas the traditional pathway included a screening based on determination of prostate serum antigen (PSA) levels and digital rectal examination followed by a systematic random transrectal biopsy\cite{Mottet2017}, in recent years the introduction of pre-biopsy mpMRI has enabled to better select patients who ought to have a prostate biopsy, increasing the diagnostic yield of the procedure\cite{Ahmed2017,Singh2018}.

To promote global standardization in the interpretation of prostate mpMRI examinations, the Prostate Imaging Reporting and Data System (PI-RADS) in its latest 2.1 version combines available evidence to assign scores to objective findings in each sequence\cite{Turkbey2019a}. However, mpMRI interpretation is time-consuming, expertise dependent\cite{Gaziev2016}, and is usually accompanied by a non-negligible inter-observer variability\cite{Sonn2019, Rosenkrantz2016}. This is particularly the case outside of expert high-volume centers\cite{Kohestani2019}. Furthermore, as with any human-based decision, MRI interpretation is not free of mistakes, which can be accentuated by cognitive impairment circumstances such as mental fatigue\cite{Lee2013}.

Computer-aided diagnosis (CAD) systems have been broadly defined as ``the use of computer algorithms to aid the image interpretation process''\cite{Giger2008}. In this sense, CAD is one of the most exciting lines of research in medical imaging and has been successfully applied to interpret images in different medical scenarios\cite{Morton2006,Gur2004,Wenzel2002,Uemura2020}. CAD poses several theoretical advantages, namely speeding up the diagnosis, reducing diagnostic errors, and improving quantitative evaluation\cite{VanGinneken2011}.

On the topic of mpMRI-based PCa CAD, different methods have been proposed since the early 2000s\cite{Chan2003,Tiwari2012,Niaf2012}. These pioneered the field but were nonetheless limited in some important aspects (e.g.: they lacked proper evaluation, expert comparison, and large enough datasets). In 2014, Litjens et al.\cite{Litjens2014a} proposed the first CAD system able to provide lesions candidate regions along with their likelihood for malignancy using pharmacokinetic, symmetry, and appearance-derived features from several MRI sequences using classical (non-Deep Learning) voxel-based classification algorithms and evaluated the results on a large cohort of 347 patients.

Since the advent of Deep Learning\cite{Krizhevsky}, however, Deep Convolutional Neural Networks (CNNs) have quickly dominated all kinds of image analysis applications (medical and otherwise), phasing out classical classification techniques. In the context of the prostate, the turning point can be traced back to the ProstateX challenge in 2016\cite{Litjens2014a,Litjens2017,Armato2018}. The challenge consisted in the classification of clinically significant PCa (csPCa) given some tentative locations in the mpMRI. More importantly, a training set of 204 mpMRIs (330 lesions) was provided openly for training the models, hence enabling many researchers up to this moment to venture into the problem (further details of this dataset can be found in Section~\ref{data-description}). At the time, half of the contestants employed classical classification methods\cite{Kitchen2017} and the other half CNNs\cite{Liu,Seah2017,Mehrtash2017}. In all cases, a patch (or region of interest, ROI) of the mpMRI around the provided lesion position was extracted, and a machine learning algorithm was trained to classify it as either csPCa or not. For instance, the second-highest-scoring method\cite{Liu}, with a lesion-wise AUC of 0.84, used a simple VGG-like\cite{Simonyan2015} CNN architecture, XmasNet, trained over the mpMRI patches to perform classification. The main limitation of all these approaches is that ROIs must be manually located beforehand, hence limiting their interest and applicability to clinical practice. Ideally, a CAD system should be able to detect all lesions' instances in a mpMRI along with their significance score.

In 2019, Yoo et al.\cite{Yoo2019} employed a ResNet\cite{He2016} to perform csPCa classification on individual mpMRI slices. Then, the probabilities of all slices were combined to produce a patient-level score. Similarly, in Xu et al.\cite{Xu2019} a modified ResNet was used to perform lesion segmentation on individual mpMRI slices, which were then combined to generate a heat map of csPCa for the whole gland. Likewise, Cao et al.\cite{Cao2019} employed a slice-wise segmentation CNN, FocalNet, not only to predict csPCa but also a map of the Gleason grade group (GGG)\cite{Epstein2005,Epstein2016} of the prostate; particularly, the model produced a lesion segmentation mask with 5 channels as output, each representing the probability of a voxel of being at least some given GGG (e.g.: GGG$\geq$1 for the first mask), with increasing GGG for every consecutive channel. Segmentation-based models are clinically interesting because they provide a csPCa map of the prostate but are otherwise cumbersome to work with since it is usually difficult (and very heuristic-reliant) to identify the lesions as individual entities and assign a score to each one, as is a common procedure in clinical practice; a similar problem arises when a patient-level score is to be generated from a lesion segmentation mask. This is natively solved in an instance detection+segmentation framework, which is very common in natural image detection tasks\cite{He2020}; yet, to the author's knowledge, no papers exist within this field employing such an approach. Additionally, 2-Dimensional (2D) slice-wise CNNs are known to generally underperform as compared with actual 3-Dimensional (3D) CNNs in lesion detection tasks\cite{Jaeger2020}.

Indeed, in 2020 several authors turned to 3D CNNs for PCa detection tasks; for instance, Aldoj et al.\cite{Aldoj2020} employed a simple 3D CNN to perform lesion classification on the ProstateX dataset and Arif et al.\cite{Arif2020} used a 3D U-Net\cite{Ronneberger2015} to perform csPCa lesion segmentation. In Schelb et al.\cite{Schelb2020}, the performance of a previously developed csPCa segmentation model was evaluated prospectively on a simulated clinical deployment over the following three years, reaching a performance comparable to expert-provided PI-RADS. Other recent papers are similar in methodology to the previous ones\cite{Nguyen2021,Riley1964,Saha2020}. Vente et al.\cite{Vente2021} provided a comprehensive analysis of what techniques are (and are not) important for performing GGG classification within a csPCa segmentation framework, and found out that soft-label ordinal regression (scaling the GGG scale system to 0 to 1 scale and performing regression) performed best. Winkel et al.\cite{Winkel2020} presented perhaps the most similar model to an actual detection framework, combining three CNNs: a Candidate Localization Network (a segmentation CNN), from which local maxima ROIs are extracted and fed into a Candidate Qualification Network that detects and eliminates false positives, leaving the rest to a final classification network, that stages the ROIs in the PI-RADS scale.

To the best of our knowledge, the model we propose is the first to leverage a proper instance detection and segmentation network, the 3D Retina U-Net\cite{Jaeger2020}, to simultaneously perform detection, segmentation, and Gleason Grade estimation from mpMRIs to a state-of-the-art performance level. It is also one of the few works that combines two very different mpMRI datasets into a single model: the ProstateX dataset and the IVO (Valencian Institute of Oncology Foundation) dataset (view Section~\ref{data-description}), achieving similarly excellent results in both. It uses prior prostate zonal segmentation information, which is provided by an automatic segmentation model, and leverages an automatic non-rigid MRI sequence registration algorithm, among other subsystems, allowing for a fully automatic system that requires no intervention. The code of this project has been made available online at \url{https://github.com/OscarPellicer/prostate_lesion_detection}.

\section{Materials and Methods}\label{materials-and-methods}
\subsection{Data description}\label{data-description}

For the development and validation of the model, two main prostate mpMRI datasets were employed: ProstateX\cite{Litjens2014a}, which is part of an ongoing online challenge at \url{https://prostatex.grand-challenge.org} and is freely available for download\cite{Litjens2017}; and IVO, from the homonymous Valencian Institute of Oncology, with permission from the Ethical Committee (CEIm-FIVO) and restricted in use to the current study.

For ProstateX, the data consisted of a total of 204 mpMRIs (one per patient) including the following sequences: T2-weighted (T2), diffusion-weighted (DW) with b-values b50, b400, and b800 s/mm\textsuperscript{2}, apparent diffusion coefficient (ADC) map (calculated from the b-values), and $K^{trans}$ (computed from dynamic contrast-enhanced -DCE- T1-weighted series). For each of these patients, one to four (1.62 per patient on average) lesion locations (i.e.: a point marking their position) and their GGG are provided (GGG is provided as part of the ProstateX2 challenge, which shares the same data with ProstateX). The lesion locations were reported by or under the supervision of an expert radiologist with more than 20 years of experience in prostate MR and confirmed by MR-guided biopsy. Furthermore, 140 additional mpMRIs are provided as part of the challenge set, including all previous information except for the GGG of the lesions. All mpMRIs were acquired by two different Siemens 3-Tesla scanners.

For IVO, there were a total of 221 mpMRIs, including the following sequences: T2, DW with b-values b100, b500, and b1000 s/mm\textsuperscript{2} (in 1.36\% of the cases, b1400 was available, instead of b1000), ADC (4.52\% missing) and a temporal series of 30 DCE T1-weighted images (42.53\% missing). For each mpMRI, one to two (1.04 per patient) lesions were segmented by one of several radiologists with two to seven years of experience in PCa imaging, and their PI-RADS were provided. The Gleason Score (GS)\cite{Epstein2005} was assessed by transperineal fusion-guided and systematic template biopsy, in which 20-30 cylinders were extracted, and two to three cylinders were directed to each of the regions of interest (ROI). Most IVO mpMRIs were acquired with a General Electric 1.5-Tesla scanner.

Four PCa classes were considered: GGG0 or benign (57.32\% of all lesions), GGG1 (GS 3+3, 17.28\%), GGG2 (GS 3+4, 12.70\%), and GGG3+ (GS 4+3, 12.70\%); therefore, lesions of GGG$\geq$3 were grouped into a single category to try to balance the classes, and also because the protocol for a suspect GGG 3+ lesion would be similar irrespective of its specific grade (i.e.: the lesion would be biopsied for confirmation). Performing accurate GGG classification from a mpMRI is a very difficult task; however, at the very least, the model should benefit from the additional information provided by the GGG label.

\subsection{Pre-processing}\label{pre-processing}

\begin{figure*}[!tb]
	\centering
	\begin{widepage}
		\includegraphics[width=0.089\linewidth]{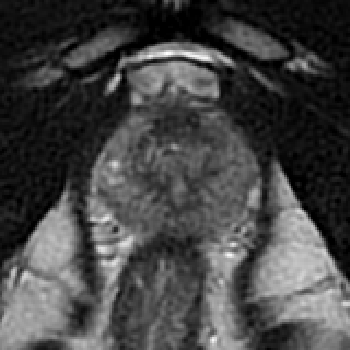}\hfill
		\includegraphics[width=0.089\linewidth]{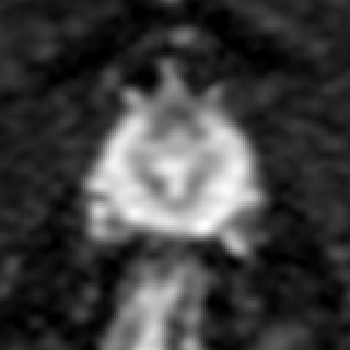}\hfill
		\includegraphics[width=0.089\linewidth]{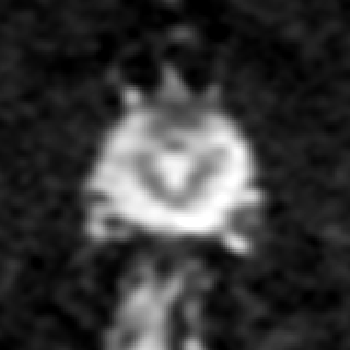}\hfill
		\includegraphics[width=0.089\linewidth]{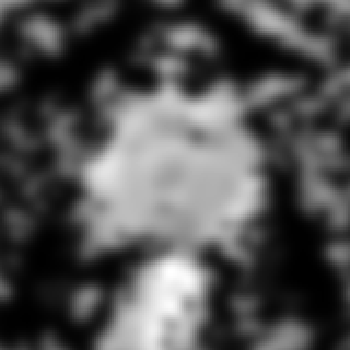}\hfill
		\includegraphics[width=0.089\linewidth]{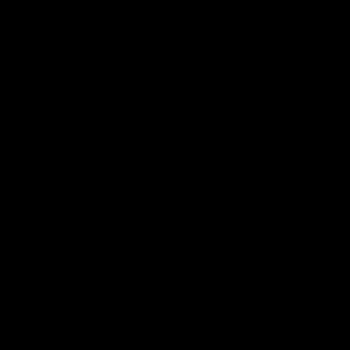}\hfill
		\includegraphics[width=0.089\linewidth]{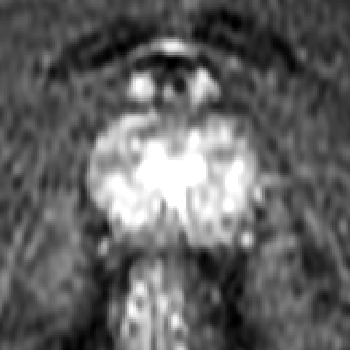}\hfill
		\includegraphics[width=0.089\linewidth]{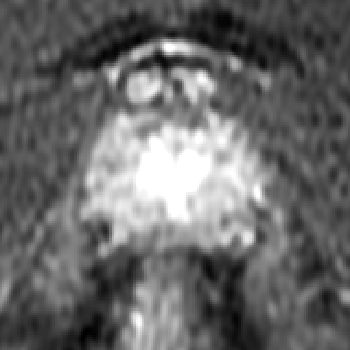}\hfill
		\includegraphics[width=0.089\linewidth]{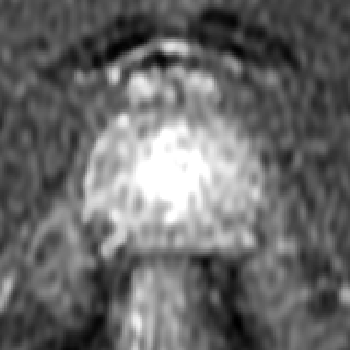}\hfill
		\includegraphics[width=0.089\linewidth]{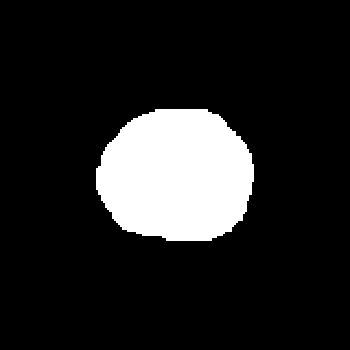}\hfill
		\includegraphics[width=0.089\linewidth]{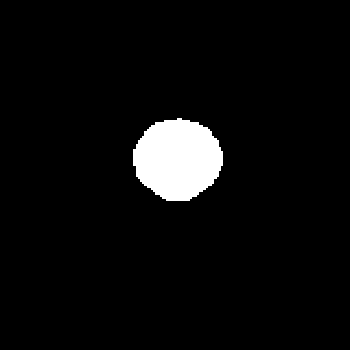}\hfill
		\includegraphics[width=0.089\linewidth]{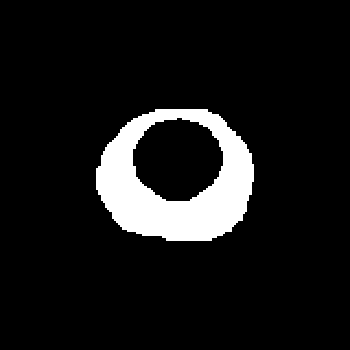}\\
		
		\vspace{0.5mm}
		
		\includegraphics[width=0.089\linewidth]{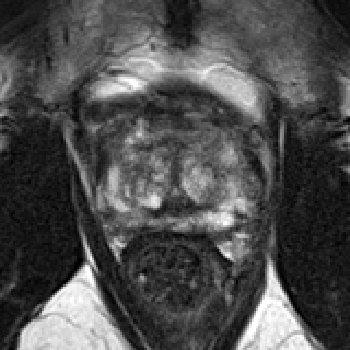}\hfill
		\includegraphics[width=0.089\linewidth]{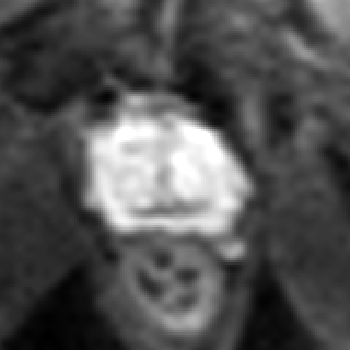}\hfill
		\includegraphics[width=0.089\linewidth]{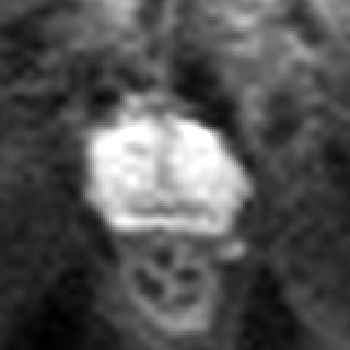}\hfill
		\includegraphics[width=0.089\linewidth]{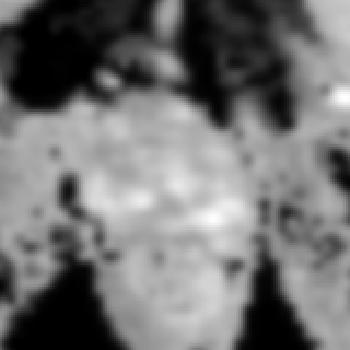}\hfill
		\includegraphics[width=0.089\linewidth]{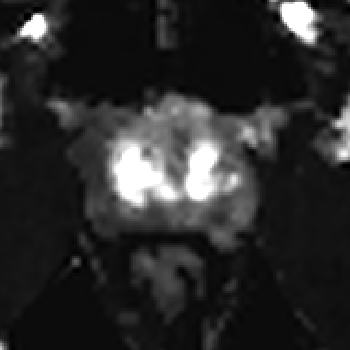}\hfill
		\includegraphics[width=0.089\linewidth]{blank.png}\hfill
		\includegraphics[width=0.089\linewidth]{blank.png}\hfill
		\includegraphics[width=0.089\linewidth]{blank.png}\hfill
		\includegraphics[width=0.089\linewidth]{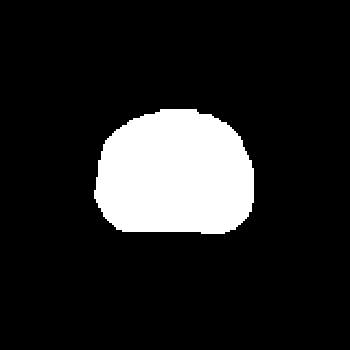}\hfill
		\includegraphics[width=0.089\linewidth]{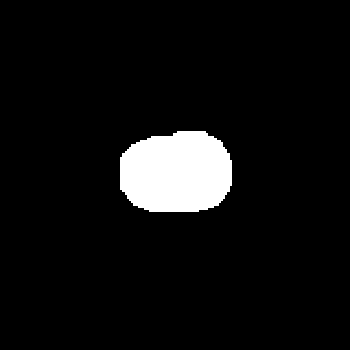}\hfill
		\includegraphics[width=0.089\linewidth]{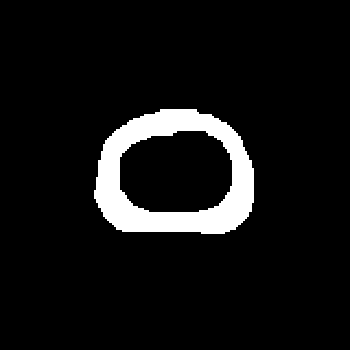}\\
	\end{widepage}
	\caption{Final pre-processed image from a single patient (top: IVO, bottom: ProstateX). Channels (from left to right): T2, b400/b500, b800/b1000/b1400, ADC, $K^{trans}$, DCE $t=10$,  DCE $t=20$, DCE $t=30$, prostate mask, CG mask and PZ mask.}\label{fig:input_images}
\end{figure*}

After collecting them, mpMRIs had to be pre-processed to accomplish three main objectives, namely: (1) homogenize differences within datasets, (2) homogenize differences between datasets, and (3) enrich the images with extra information that might be useful for the model. Additionally, the preprocessing pipeline was designed to require as little human intervention as possible, in pursuit of developing a system easily implementable in clinical practice.

For the first objective, all images were cropped to an ROI around the prostate of size $160 \times 160 \times 24$ voxels with a spacing of $\left( 0.5,\ 0.5,\ 3 \right)$mm, which corresponds with the median (and mode) spacing of the T2 sequences for both datasets. The rest of the sequences were applied the same processing for the sake of homogeneity. B-Spline interpolation of third order was employed for all image interpolation tasks, while Gaussian label interpolation was used for the segmentation masks. For the IVO dataset, the time series of 30 DCE images per patient was sampled at times 10, 20, and 30, approximately coinciding with the peak, progression, and decay of the contrast agent. Then, all sequences were combined into a single multi-channel image, in which any missing sequences were left blanks (value of 0), such as the three DCE channels in every ProstateX image, or the $K^{trans}$ channel in every IVO image. The intensity was normalized by applying Equation~\ref{eq:normalization} to every channel of an image \emph{I} independently, as introduced in Pellicer-Valero et al.\cite{Pellicer-valero2021}.

\begin{equation}\label{eq:normalization}
	I_{new}=\frac{I - percentile(I, 1)}{percentile(I, 99) - percentile(I, 1)}
\end{equation}

Regarding objective (2), the procedure for homogenizing lesion representations between datasets is described in Section~\ref{automated-lesion-growing}, and a special data augmentation employed to alleviate the problem of missing sequences is presented in Section~\ref{online-data-augmentation}. Additionally, sequences b500 (from IVO) and b400 (from ProstateX) were considered similar enough to conform to the same channel in the final image; likewise, sequences b1000/b1400 (from IVO) and b800 (from ProstateX) were assigned to a single common channel too.

Concerning objective (3), Section~\ref{automated-prostate-zonal-segmentation} argues that prostate zonal segmentation is an important input for PCa assessment and describes the conception of a model for producing such segmentations automatically. Finally, DW and ADC sequences were found to be misaligned to the rest of the sequences in several patients; hence an automated registration step was added, which is presented in Section~\ref{automated-sequence-registration}.

Figure~\ref{fig:input_images} shows the channels of one image from each dataset after all the mentioned pre-processing steps. 

\subsubsection{Automated lesion growing}\label{automated-lesion-growing}

To enable training a single model on both datasets, it was mandatory to homogenize how lesion information was to be provided to the model: while the IVO dataset provided the full segmentation mask for each lesion, in ProstateX only the center position of the lesion was available. Although detection systems can be adapted to detect positions, they are typically designed to work with much more semantically rich bounding boxes (BB)~\cite{He2020}, or segmentations, or both~\cite{Ren2017}.

\begin{figure}[h]
	\centering
	\includegraphics[width=0.33\linewidth]{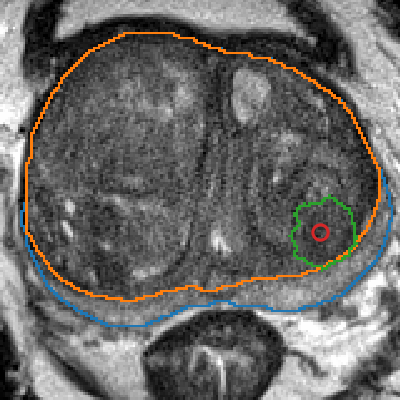}\hfill
	\includegraphics[width=0.33\linewidth]{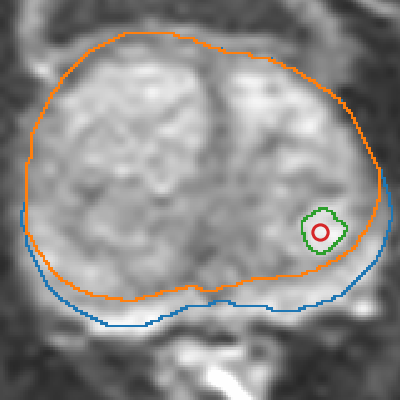}\hfill
	\includegraphics[width=0.33\linewidth]{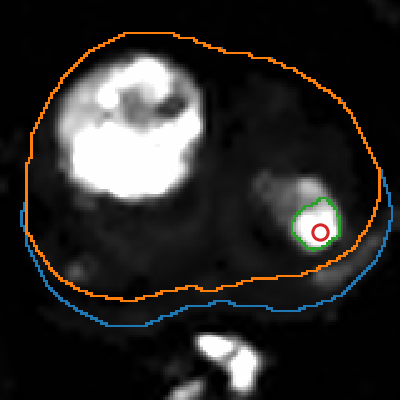}
	\caption{Automatic lesion segmentation for a ProstateX patient in sequences (from left to right: T2, b800 and $K^{trans}$) before combining them. Prostate zonal segmentation and the original lesion position (in red) are shown for reference.}\label{fig:lesion_growing}
\end{figure}

\begin{figure*}[!t]
	\centering
	\begin{widepage}
		\includegraphics[width=\textwidth]{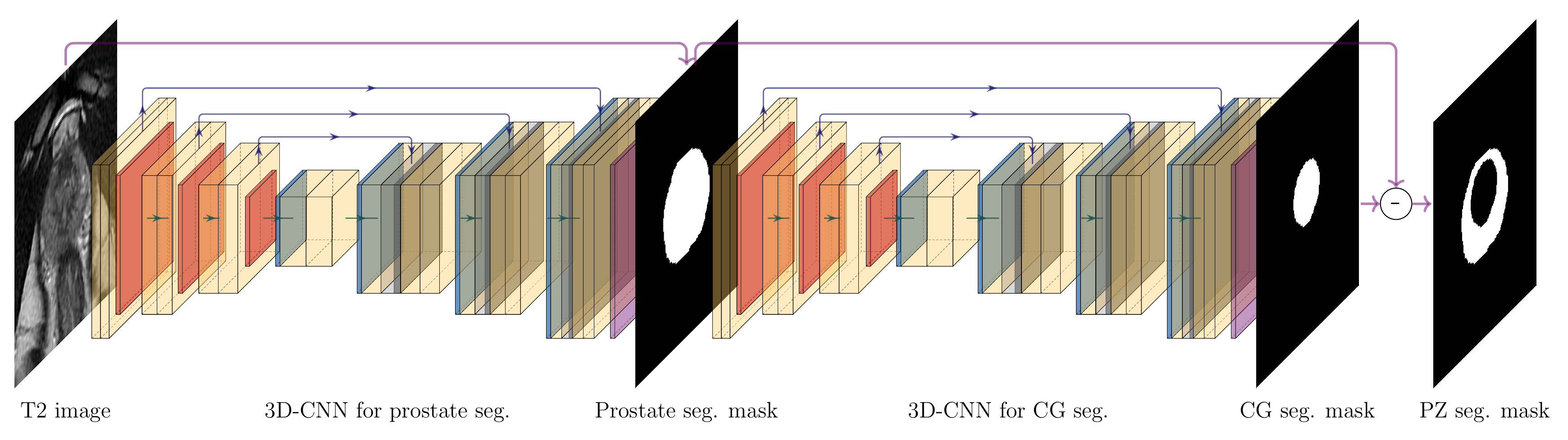}
	\end{widepage}
	\caption{Cascading CNNs for prostate CG and PZ segmentation. The first 3D-CNN takes a T2 image as input and produces a prostate segmentation mask as output, while the second 3D-CNN takes both the T2 image and the prostate segmentation as inputs to produce the CG segmentation mask as output. Finally, PZ is computed by subtraction of both output masks.}\label{fig:cascade}
\end{figure*}  

To solve this inconsistency between the datasets, a similar approach to Liu et al.\cite{Liu2019} was employed: for the ProstateX dataset, lesions were automatically segmented by growing them from the provided image position (used as seed), using a threshold level set method from Python library SimpleITK\cite{Yaniv2018}. Concretely, the algorithm was applied independently to sequences T2, b800, and $K^{trans}$, and all segmented areas present in at least two of these three sequences were kept. Figure~\ref{fig:lesion_growing} shows the process of applying this segmentation algorithm to one image.

\subsubsection{Automated prostate zonal segmentation}\label{automated-prostate-zonal-segmentation}

Following McNeal's criterion\cite{Selman2011}, the prostate is typically partitioned into two distinct zones: the Central Gland (CG, including both the transition zone and the central zone, which are difficult to distinguish) and the Peripheral Zone (PZ). PCa lesions vary in frequency and malignancy depending on zone\cite{Haffner2009} and, as such, PI-RADS v2 considers them when assessing mpMRIs\cite{Weinreb2016}. Therefore, just like a radiologist, a model for automated PCa detection and classification will probably also benefit from having both CG and PZ mask priors provided as inputs, in addition to the mpMRI.

Accordingly, a cascading system of two segmentation CNNs, similar to the one introduced by Zhu et al.\cite{Zhu2019}, was developed for automatic CG and PZ segmentation. As it can be seen in Figure~\ref{fig:cascade}, the first CNN (which is a published model\cite{Pellicer-valero2021} based on the U-Net(Ronneberger, Fischer, and Brox 2015) CNN architecture with dense\cite{Huang2017} and residual\cite{He2016} blocks) takes a prostate T2 image as input and produces a prostate segmentation mask as output. Then, the second CNN takes both the T2 image and the prostate segmentation mask obtained in the previous step and generates a CG segmentation mask as output. Finally, the PZ segmentation mask can be computed by subtracting the CG from the prostate segmentation mask.

The second CNN employed an architecture identical to the first one but was retrained on 92 prostate T2 images from a private dataset, in which the CG was manually segmented by a radiologist with two years of experience in PCa imaging. To be more precise, 80 of the 92 images were used for training the CG segmentation model, while the remaining 12 were employed for testing. Additionally, this model was also blindly tested (i.e.: with no retraining or adaptation of any kind) against the NCI-ISBI\cite{N2015} train dataset, which is freely available at \url{http://doi.org/10.7937/K9/TCIA.2015.zF0vlOPv} The results of this prostate zonal segmentation model are very briefly analyzed and compared to others in Section~\ref{prostate-zonal-segmentation} Once trained and validated, this model was employed to obtain the CG and PZ masks of all the prostates in the current study.

For further information on this topic, please see, for instance: Liu et al.\cite{Liu2020}, where the authors include Bayesian uncertainty to a zonal segmentation CNN; Qin et al.\cite{Qin2020}, where a multi-directional edge loss is employed; Chilali et al.\cite{Chilali2016}, in which a more classical approach with C-means clustering is utilized; Rundo et al.\cite{Rundo2019}, where Squeeze-and-Excite\cite{Hu2018} blocks are added to the CNN architecture; or Can et al.\cite{Can2018}, in which only scribbles are provided as Ground Truth (GT) for training the CNN.

\subsubsection{Automated sequence registration}\label{automated-sequence-registration}

\begin{figure}[h]
	\centering
	\includegraphics[width=0.33\linewidth]{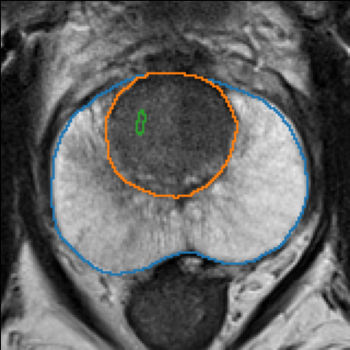}\hfill
	\includegraphics[width=0.33\linewidth]{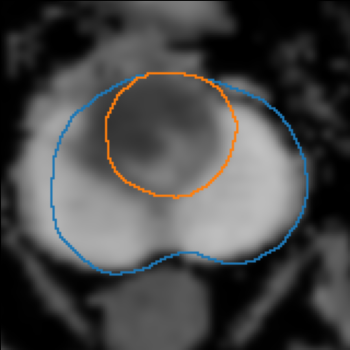}\hfill
	\includegraphics[width=0.33\linewidth]{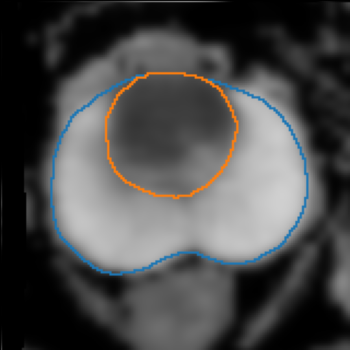}
	\caption{Automatic registration between T2 sequence (left) and ADC map (center: before, right: after) for a sample mpMRI.}\label{fig:registration}
\end{figure}

In several patients, DW sequences and the ADC map were misaligned to T2 and the other sequences. As a solution, non-rigid registration (based on a BSpline transformation) was applied between the spatial gradient of the T2 and the ADC map using Python library SimpleITK\cite{Yaniv2018}, with Mattes Mutual Information\cite{Mattes2001} as loss function and gradient descent\cite{Ruder} as the optimizer for the BSpline parameters. For every mpMRI, the registration algorithm was run 50 times with different parameter initializations, and the correlation coefficient between the spatial gradient of the T2 sequence and the spatial gradient of the registered ADC map was evaluated at the CG and the PZ areas. These custom metrics allowed to place a bigger emphasis to the areas of interest, as compared to image-wide metrics. Finally, the transformation associated with the run yielding the highest value for the average of all metrics and the loss was chosen as final and applied to both DW and ADC sequences. Figure~\ref{fig:registration} shows the result of applying this procedure to one mpMRI.

\subsection{Model training and validation}\label{model-training-and-validation}

After pre-processing the data, it was used to train a Retina U-Net\cite{Jaeger2020} CNN architecture, which allows for the simultaneous detection, segmentation, and classification of PCa lesions. Section~\ref{architecture-retina-u-net} provides an overview of this architecture, while Sections~\ref{hyperparameters}-\ref{epoch-and-cv-ensembling-during-testing} deal with all engineering decisions related to the model training, validation, and testing.

\subsubsection{Architecture: Retina U-Net}\label{architecture-retina-u-net}

The Retina U-Net architecture combines the Retina Net\cite{Lin2017} detector with the U-Net segmentation CNN(Ronneberger, Fischer, and Brox 2015) and is specifically designed for application to medical images. On one hand, Retina Net is a one-shot detector, meaning that classification and BB refinement (regression) are directly performed using the intermediate activation maps from the output of each decoder block in the Feature Pyramid Network (FPN) that conforms its backbone\cite{Lin2017a}, making it not only more efficient but also better suited for lesion detection in medical images, which have distinct characteristics compared to natural images (e.g.: there is no overlap between detections).

\begin{figure}[h]
	\centering
	\includegraphics[width=\linewidth]{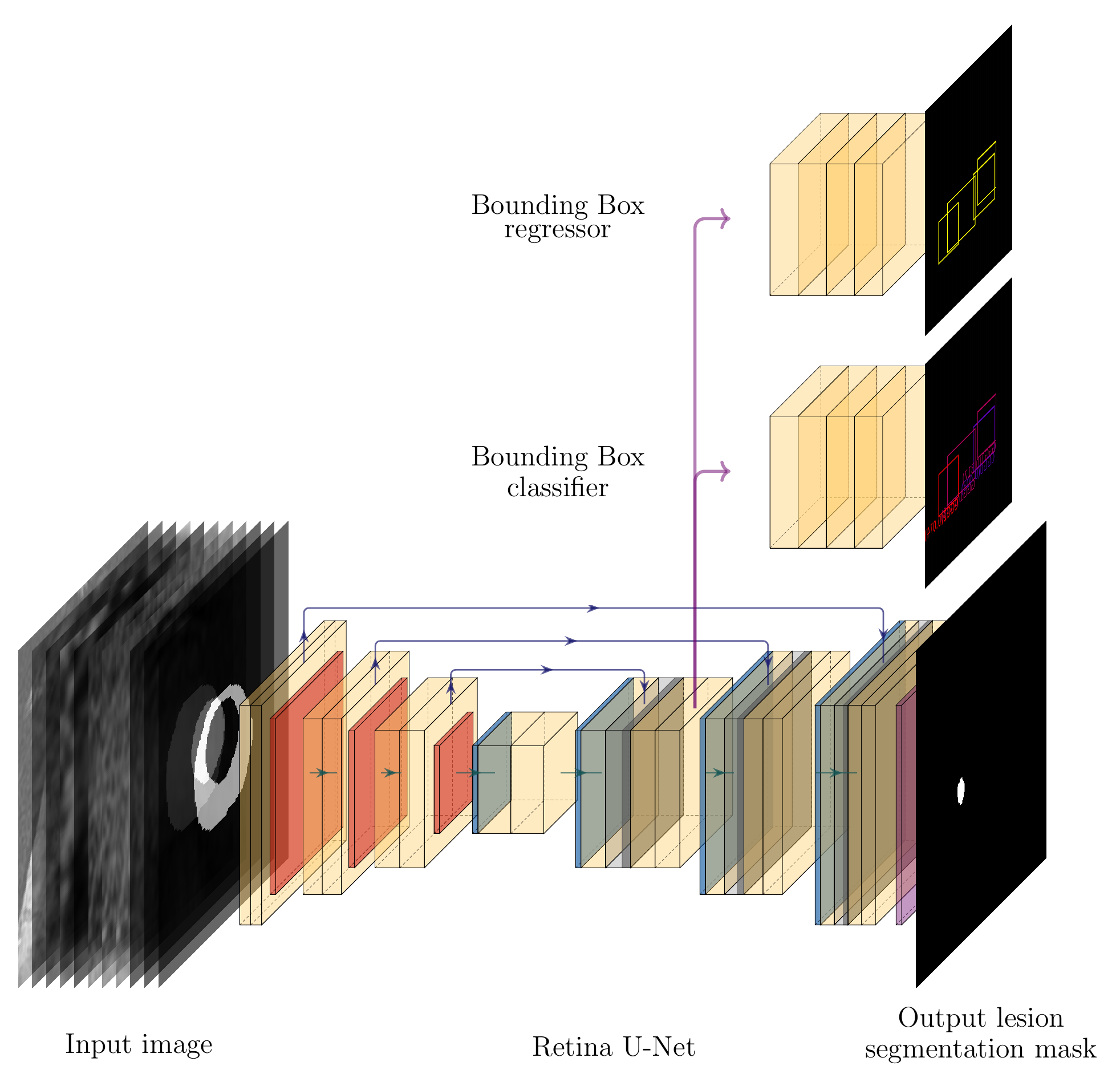}
	\caption{Overview of the Retina U-Net architecture. On the bottom, a U-Net-like architecture segments the lesions present in the image irrespective of their class. On the top, a BB regression head takes a feature map from a decoder of the U-Net and refines the coarse detections, while the BB classifier tries to predict their class. These two heads visit all decoder levels, performing detection at different scales transparently.}\label{fig:retina unet}
\end{figure}

Furthermore, in the Retina U-Net, the FPN has been extended with two more high-resolution pyramid levels leading to a final segmentation layer, hence making the extended FPN architecture extremely akin to that of the U-Net. Therefore, the lesions are segmented independently of the detections (unlike other similar detection+segmentation architectures, such as Mask R-CNN\cite{He2020}). This simplifies the architecture significantly, while still being a sensible choice for segmenting lesions since they all represent a single entity irrespective of their particular classes. Figure~\ref{fig:retina unet} shows an overview of the Retina U-net architecture applied to the problem of simultaneous PCa detection, classification, and segmentation.

\subsubsection{Hyperparameters}\label{hyperparameters}

An ensemble of five CNNs (see Section~\ref{epoch-and-cv-ensembling-during-testing}) was trained with the ResNet101 backbone\cite{He2016} with batch normalization\cite{Ioffe2015} and a batch size of 6, at 120 batches per epoch, for a total of 115 epochs. Please, refer to Section~\ref{data-partitioning} for more information on how data was split for training and validating the model. A triangular cyclical Learning Rate (LR) with exponential decay was employed\cite{Smith2017}, with LRs oscillating between a minimum of  $8\cdot10^{-5}$ and a maximum of  $3.5\cdot10^{-4}$. For the BBs, a single aspect ratio of 1 (before BB refinement) was considered sufficient, with scales ranging from $4 \times 4 \times 1$ voxels (i.e.: $2 \times 2\  \times 3$ mm), all the way to $28 \times 28 \times 9$ voxels (i.e.: $14 \times 14 \times 27$ mm), depending on the pyramid level on which the detection was performed. The rest of the parameters were left at the default values according to the original publication\cite{Jaeger2020}.

\subsubsection{Online data augmentation}\label{online-data-augmentation}

To help with regularization, and to expand the limited training data, extensive online 3D data augmentation was employed during training using the Python library Batchgenerators\cite{Fabian2020}, including both rigid and non-rigid transformations, such as scaling, rotations, and elastic deformations.

Additionally, a custom augmentation was included to help deal with the issue of missing sequences, either because they never existed (such as $K^{trans}$ images in the IVO dataset), or because they were missing. This augmentation, named Random Channel Drop, consists in -as its name suggests- setting any given channel to zero (blanking it) with a certain probability, hence accustoming the model to dealing with missing data. Therefore, during training, every channel of every image had a 7.5\% probability of being dropped at any time, except for the T2 channel and the segmentation masks, which had a probability of 0\%, since they are assumed to be always available. The three DCE channels were considered as a whole for the purposes of dropping them (i.e.: they could not be dropped independently of each other).

\subsubsection{Data partitioning}\label{data-partitioning}

The mpMRIs were split into two sets: the train/validation set and the test set. The test set only contained ``complete'' mpMRIs (with no missing sequences), amounting to 30 IVO patients (23.62\% of all complete IVO patients) and 45 ProstateX patients (22.17\% of all ProstateX patients). This set was kept secret during the development of the model and was only employed eventually to validate it. Instead, for internal validation, five-fold cross-validation (CV) was employed: the train/validation set was split into five disjoint subsets, and five different instances of the same Retina U-Net model were successively trained on four out of the five subsets and validated on the fifth, hence creating a virtual validation dataset that encompassed the totality of the training data (but not the test data, which were kept apart).

As mentioned in Section~\ref{data-description}, there was an additional ProstateX challenge set containing 140 mpMRIs with all the same information as the training set, except for the lesion GGG, which was not available. Hence, this dataset could also be employed for training both the segmentation and the BB regressor components of the Retina U-Net (but not the classifier). As such, this dataset was included as part of the training set (but not in the validation sets, as it contained no GT class information), and the classifier had to be modified to ignore any detection belonging to this dataset (i.e.: the loss was not propagated from such detections).

In summary, the model was trained and five-fold cross-validated with 191 IVO patients (of which only 45.55\% were complete) + 159 ProstateX patients (all complete) + 140 ProstateX test patients (those coming from the ProstateX challenge set, for which GGG class information was not available). For testing, a secret subset consisting of 30 IVO patients and 45 ProstateX patients (all complete) was employed. The model was also tested on the ongoing ProstateX challenge.

\subsubsection{Epoch and CV ensembling during testing}\label{epoch-and-cv-ensembling-during-testing}

During the final test set prediction, both epoch and CV ensembling were used to boost the capabilities of the model. In general, ensembling consists in training \emph{N} models for the same task, using them to predict on a given test set, and then somehow combining all \emph{N} predictions to achieve a better joint performance than that of each model individually.

Hence, the five CV models were used for ensembling and, additionally, for every one of these CV models, each of the best five was used as a further independent model, hence totaling an equivalent of 25 virtual models.

Then, the predictions from the ensemble on the test set were combined in the following way: for segmentation masks, the average mask was obtained and, for the BBs, the weighted box clustering (WBC) algorithm with an Intersection over Union (view Section~\ref{metrics}) threshold of $1\cdot10^{-5}$ was applied to each class independently. The WBC algorithm is described in the original Retina U-Net paper\cite{Jaeger2020}.

\subsection{Metrics}\label{metrics}

This section introduces the metrics that will be employed for assessing the results. In general, and unless otherwise noted, metrics are a number (in the range from 0 to 1, with 1 representing a perfect score) that compares some GT data with some predicted data:

\begin{itemize}
\item
Sørensen-Dice similarity coefficient (DSC): Used for assessing the overlap between two segmentations, it is calculated as two times the volume of their intersection divided by the sum of their volumes. It is usually computed patient-wise and aggregated for all patients using the mean.
\item
Intersection over Union (IoU): Commonly employed for evaluating the overlap between two BBs, it is calculated as the volume of intersection between two BBs divided by the volume of their union.
\item
Sensitivity and specificity: For binary classification (e.g.: benign vs. csPCa) sensitivity measures the ratio of clinically significant lesions that were found. E.g.: a sensitivity of 0.75 means that 75\% of all clinically significant lesions were found, yet 25\% of them remained undetected. Conversely, specificity measures the ratio of non-clinically significant lesions that were found. E.g.: a specificity of 0.75 means that 75\% of all non-clinically significant lesions have been correctly identified, while 25\% have been misclassified as being significant. Classification models typically associate a confidence score to each prediction (i.e.: a value from 0 to 1), which can be used to tune the sensitivity-specificity interrelationship at will, by adjusting the confidence threshold. E.g.: by setting the threshold to 0.01, almost all lesions will be considered clinically significant, leading to very high sensitivity yet very low specificity; while, by setting the threshold to 0.99, the opposite would happen.
\item
The area under the receiver operating characteristic curve (AUC-ROC, or just AUC): The AUC summarizes the sensitivity and specificity for all possible threshold values and is computed as the area under the curve that emerges by plotting the sensitivity against 1-specificity, for all possible threshold values.
\end{itemize}

\subsection{Lesion matching and evaluation}\label{lesion-matching-and-evaluation}

The results were evaluated at three lesion significance thresholds (GGG$\geq$1, GGG$\geq$2, and GGG$\geq$3) and two levels: lesion-level and patient-level. Only predicted BBs with a predicted GGG equal or above the chosen significance threshold (e.g.: GGG$\geq$2) were considered, and the rest were completely ignored.

For lesion-level evaluation, each of the GT lesions was first matched with one (or none) of the detected lesions. First, all predicted BBs whose centroid was less than 15mm away from that of the GT BB were selected as candidates for matching, and assigned a matching score computed as \(\widehat{p} + k \cdot (1 - d/15mm)\), where \(\widehat{p}\) represents the actual score given by the model to that detection, \(d\) is the distance between the GT BB centroid and the candidate BB centroid, and \(k = 2\). That way, both the model confidence (\(\widehat{p}\)) and distance to the GT (\(d\)) were considered for matching. The parameters for this matching procedure (e.g.: \(k = 2\), 15mm) were adjusted directly on the training set. If no detections existed within a 15mm radius of a GT BB, a score of 0 was assigned to it. This evaluation method measures the performance of the model only on GT lesions for which biopsy confirmation and GGG are available, without assuming anything about the rest of the prostate, which may or may not contain other lesions. Furthermore, it allows the model to compete in the online ProstateX challenge (despite it not being an ROI classification model) since it can assign a score to every GT lesion.

For patient-level evaluation, the patient score was computed as the highest score from any BB predicted for the patient, and the GT GGG of a patient was computed as the highest GGG among all his GT lesions and among all the 20-30 cylinders obtained in the systematic biopsy (which were only available for patients from the IVO dataset). Hence, for the IVO dataset, a patient without any significant GT lesions might still have csPCa; for ProstateX, however, we do not know, and we must assume that this does not happen.

\section{Results}\label{results}

\begin{table*}[t]
	\renewcommand{\arraystretch}{1.1}
	\caption{Quantitative results for ProstateX test data evaluated with different GGG significance criteria (e.g.: lesions with GGG$\geq$1,2, or 3 are considered positive), at lesion- and patient-level (\(N_{\text{positives}}\) / \(N_{\text{total}}\)), and at two thresholds ($t$): maximum sensitivity and balanced.}\label{tab:results_prostatex}
	\small
	\begin{widepage}
		\centering
		\begin{tabular}{lllllllll}
			\toprule
			\thead{\multirow{2.5}{*}{\makecell{Significance\\criterion}}} & \thead{\multirow{2.5}{*}{Level}} & \thead{\multirow{2.5}{*}{AUC}} 
			& \multicolumn{3}{c}{\thead{Max. sensitivity}} & \multicolumn{3}{c}{\thead{Balanced}}\\
			\cmidrule(lr){4-6}
			\cmidrule(lr){7-9}
			& & & \thead{$t$} & \thead{Sens.} & \thead{Spec.} & \thead{$t$} & \thead{Sens.} & \thead{Spec.}\\
			\midrule
			GGG$\geq$1 & Lesion (17/69) & 0.898 & 0.028 & 0.941 & 0.788 & 0.053 & 0.824 & 0.865\\
			& Patient (16/45) & 0.866 & 0.108 & 1.000 & 0.138 & 0.108 & 0.938 & 0.655\\
			\textbf{GGG$\geq$2} & Lesion (13/69) & 0.959 & 0.028 & 1.000 & 0.786 & 0.108 & 0.923 & 0.911\\
			& Patient (13/45) & 0.865 & 0.028 & 1.000 & 0.375 & 0.108 & 0.923 & 0.688\\
			GGG$\geq$3 & Lesion (7/69) & 0.751 & 0.195 & 0.714 & 0.887 & 0.195 & 0.714 & 0.887\\
			& Patient (7/45) & 0.767 & 0.016 & 1.000 & 0.395 & 0.026 & 0.857 & 0.500\\
			\bottomrule
		\end{tabular}
	\end{widepage}
\end{table*}

\begin{table*}[t]
	\renewcommand{\arraystretch}{1.1}
	\caption{Quantitative results for IVO test data evaluated with different GGG significance criteria (e.g.: lesions with GGG$\geq$1,2, or 3 are considered positive), at lesion- and patient-level ($N_{positives}/N_{total}$), and at two thresholds ($t$): maximum sensitivity and balanced. Results are compared with radiologist-assigned pre-biopsy PI-RADS scores for all IVO patients with no missing sequences and with PI-RADS information available (N=106 patients, 111 lesions).}\label{tab:results_ivo}
	\small
	\begin{widepage}
		\centering
		\begin{tabular}{lllllllllllll}
			\toprule
			\thead{\multirow{2.5}{*}{\makecell{Significance\\criterion}}} 
			& \thead{\multirow{2.5}{*}{Level}} & \thead{\multirow{2.5}{*}{AUC}} 
			& \multicolumn{3}{c}{\thead{Max. sensitivity}} & \multicolumn{3}{c}{\thead{Balanced}} 
			& \multicolumn{2}{c}{\thead{PI-RADS$\geq$4}} & \multicolumn{2}{c}{\thead{PI-RADS=5}}\\
			\cmidrule(lr){4-6}
			\cmidrule(lr){7-9}
			\cmidrule(lr){10-11}
			\cmidrule(lr){12-13}
			& & & \thead{$t$} & \thead{Sens.} & \thead{Spec.} & \thead{$t$} & \thead{Sens.} & \thead{Spec.} & \thead{Sens.} & \thead{Spec.} & \thead{Sens.} & \thead{Spec.}\\
			\midrule
			GGG$\geq$1 & Lesion (\(13/33\)) & 0.892 & 0.027 & 1.000 & 0.350 & 0.105 & 0.923 & 0.700 & 0.741 & 0.604 & 0.328 & 0.962\\
			& Patient (15/30) & 0.920 & 0.253 & 1.000 & 0.667 & 0.301 & 0.867 & 0.800 & 0.710 & 0.649 & 0.290 & 0.973\\
			\textbf{GGG$\geq$2} & Lesion (8/33) & 0.945 & 0.173 & 1.000 & 0.800 & 0.301 & 0.875 & 0.920 & 0.882 & 0.558 & 0.441 & 0.922\\
			& Patient (9/30) & 0.910 & 0.219 & 1.000 & 0.762 & 0.263 & 0.889 & 0.810 & 0.850 & 0.576 & 0.400 & 0.924\\
			GGG$\geq$3 & Lesion (3/33) & 0.840 & 0.301 & 1.000 & 0.778 & 0.316 & 0.667 & 0.833 & 0.727 & 0.440 & 0.455 & 0.840\\
			& Patient (3/30) & 0.840 & 0.301 & 1.000 & 0.778 & 0.316 & 0.667 & 0.815 & 0.727 & 0.432 & 0.455 & 0.832\\
			\bottomrule
		\end{tabular}
	\end{widepage}
\end{table*}

\subsection{Prostate zonal segmentation}\label{prostate-zonal-segmentation}

Regarding the prostate zonal segmentation model, which was developed with the sole purpose of automating the PCa detection system, the results for all datasets can be found in Table~\ref{tab:results_segmentation}, the DSC ranged from 0.894 to 0.941. Some qualitative results for this segmentation model can be seen in \mbox{Figures~\ref{fig:lesion_growing},~\ref{fig:registration},~\ref{fig:model_output}, and~\ref{fig:model_output2}}.

\begin{table}[h]
	\renewcommand{\arraystretch}{1.1}
	\caption{Results for prostate zonal segmentation}\label{tab:results_segmentation}
	\small
	\centering
	\begin{tabular}{lllll}
		\toprule
		\thead{\multirow{2.5}{*}{Dataset}} 
		& \thead{\multirow{2.5}{*}{N}}  
		&\multicolumn{3}{c}{\thead{Mean DSC}} \\ 
		\cmidrule(lr){3-5}
		& & 
		\thead{Prost.} & \thead{CG} & \thead{PZ}  \\
		
		\midrule			
		Private train & 80 
		& 0.941 & 0.935 & 0.866 \\
		
		Private test & 12 
		& 0.915 & 0.915 & 0.833 \\
		
		NCI-ISBI train & 60 
		& 0.894 & 0.860 & 0.690 \\
		\bottomrule
	\end{tabular}
\end{table}

\subsection{Lesion detection, segmentation, and classification}\label{lesion-detection-segmentation-and-classification}
\subsubsection{Quantitative results}\label{quantitative-results}

\begin{figure*}[tb]
	\centering
	\begin{widepage}
		\includegraphics[width=0.499\linewidth]{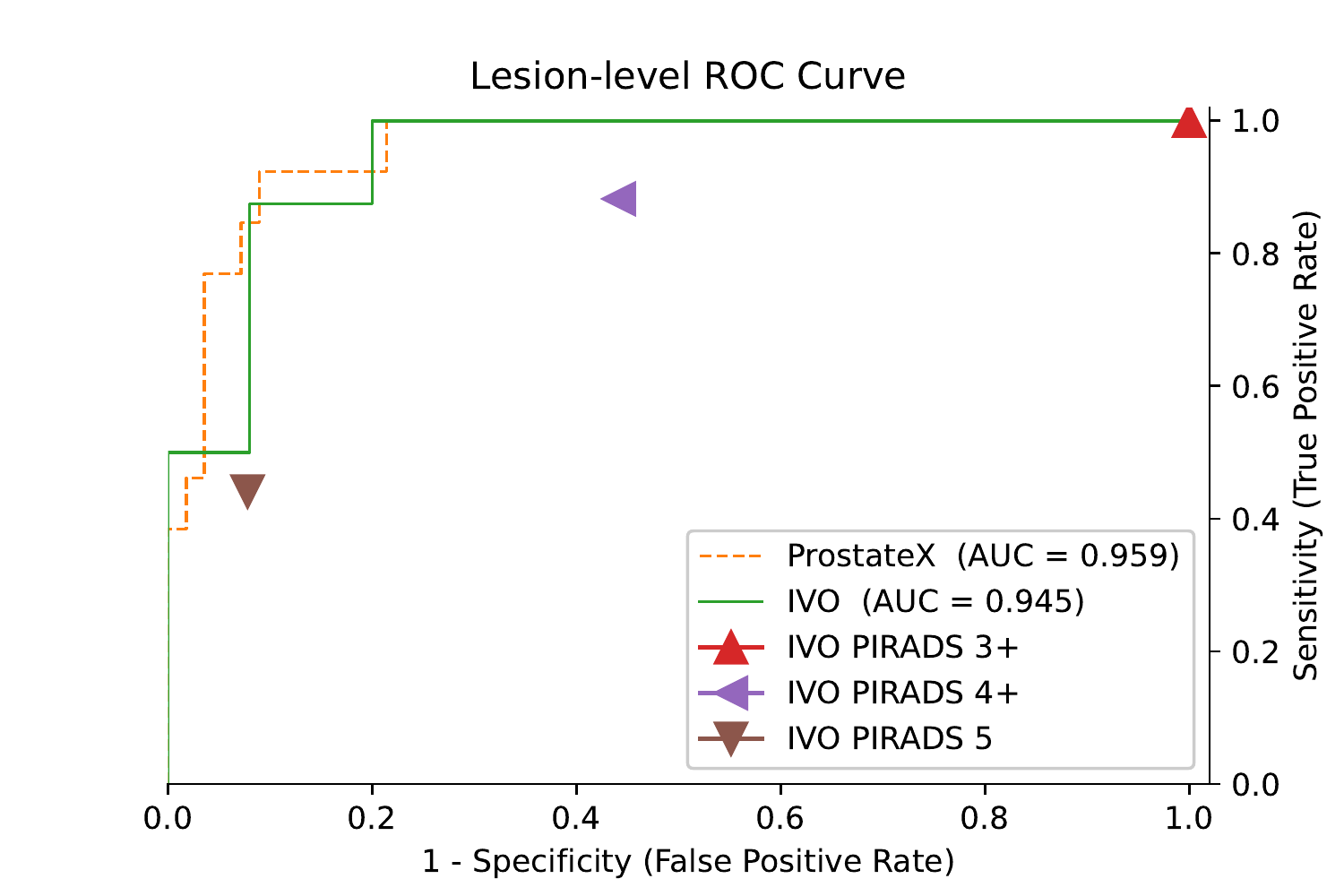}\hfill
		\includegraphics[width=0.499\linewidth]{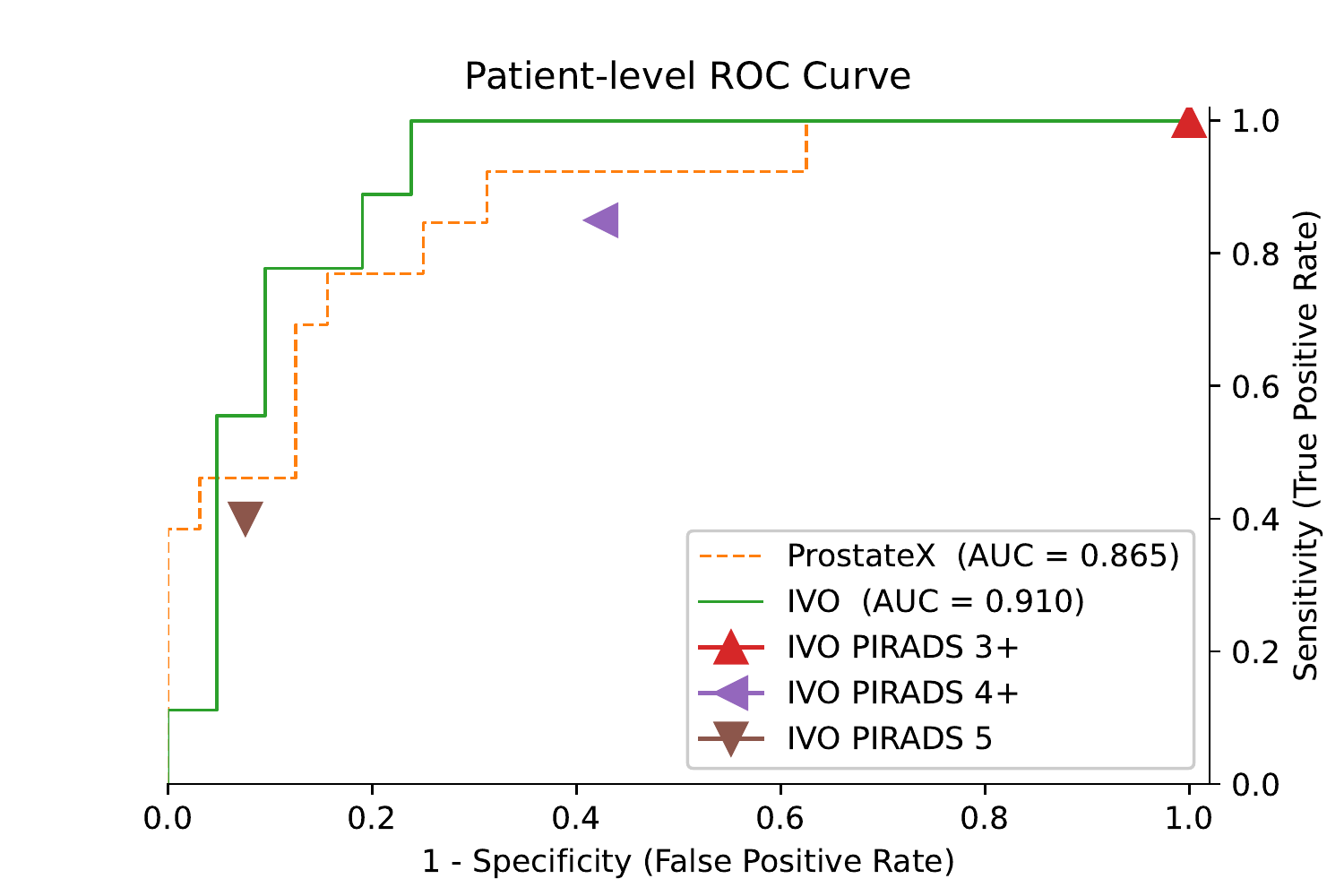}\hfill
		
	\end{widepage}
	\caption{ROC curve of the model for significance criterion GGG$\geq$2, evaluated at the lesion level (left) and the patient level (right). For comparison, triangular marks represent the radiologist-assigned pre-biopsy PI-RADS.}\label{fig:auc}
\end{figure*}

A comprehensive quantitative evaluation of the trained model on the ProstateX and IVO test sets has been compiled in Table~\ref{tab:results_prostatex} and Table~\ref{tab:results_ivo}, respectively. The employed metrics were introduced in Section~\ref{metrics} and their computation procedure was explained in Section~\ref{lesion-matching-and-evaluation}. For the evaluation of sensitivity and specificity, the model-predicted scores were thresholded at two working points (computed a posteriori on the test data): maximum sensitivity and balanced (similar sensitivity and specificity). Furthermore, radiologist-assigned pre-biopsy PI-RADS scores for all IVO patients with no missing sequences and with PI-RADS information available (N=106 patients, 111 lesions) has also been included in Table 3 for comparison. Please notice that PI-RADS$\geq 3$ is omitted since all IVO lesions were assigned at least a PI-RADS 3 score, and hence PI-RADS$\geq 3$ acts just as a naïve classifier that considers all samples as positive (sensitivity 1 and specificity 0). Finally, a graphical representation of the ROC curve for the main significance criterion (GGG$\geq$2) can be found in Figure~\ref{fig:auc}.

Focusing on the results for the GGG$\geq$2 significance criterion, at the highest sensitivity working point, the model achieves a perfect lesion-level sensitivity of 1 (no csPCa is missed) and a specificity of 0.786 and 0.875 for ProstateX and IVO, respectively (AUCs: 0.959 and 0.945). At the patient level, the specificity falls to 0.375 and 0.762 for each dataset (AUCs: 0.865 and 0.910).

For the GGG$\geq$1 significance criterion, the model achieves a lesion-/patient-level maximum sensitivity of 0.941 (spec. 0.788) / 1 (spec. 0.138) in the ProstateX dataset, and a maximum sensitivity of 1 (spec. 0.350) / 1 (spec. 0.667) in the IVO dataset. In summary, no GGG$\geq$1 patient was missed, although at a cost of low specificity. Finally, using the GGG$\geq$3 significance criterion the model reaches a lesion- and patient-level sensitivity of 0.714 (spec. 0.887) / 1 (spec.: 0.395) in the ProstateX dataset, and a maximum sensitivity of 1 (spec. 0.778) / 1 (spec. 0.778) in the IVO dataset.

Regarding lesion segmentation performance, the mean DSC across all patients for segmenting any type of lesion irrespective of their GGG (including GGG0 benign lesions), was 0.276/0.255 for the IVO/ProstateX dataset when evaluated at the 0.25 segmentation threshold, and 0.245/0.244 when evaluated at 0.5.

\hypertarget{qualitative-results}{%
\subsubsection{Qualitative results}\label{qualitative-results}}

\begin{figure*}[tb]
	\centering
	\begin{widepage}
		\includegraphics[width=0.198\linewidth]{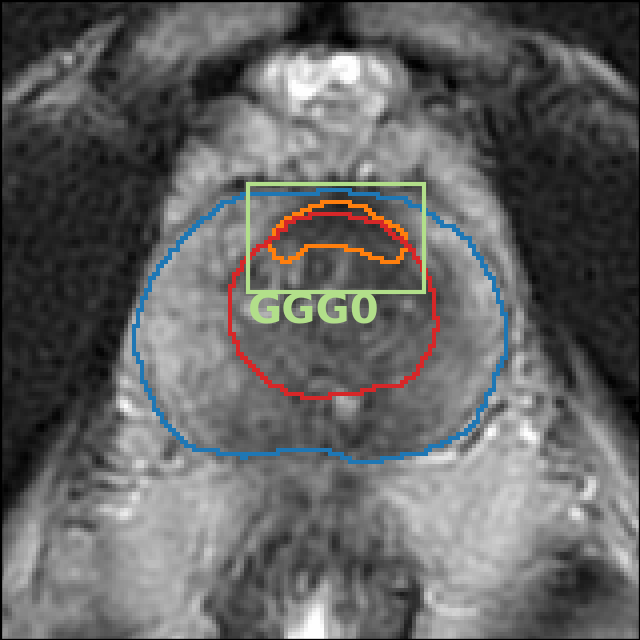}\hfill
		\includegraphics[width=0.198\linewidth]{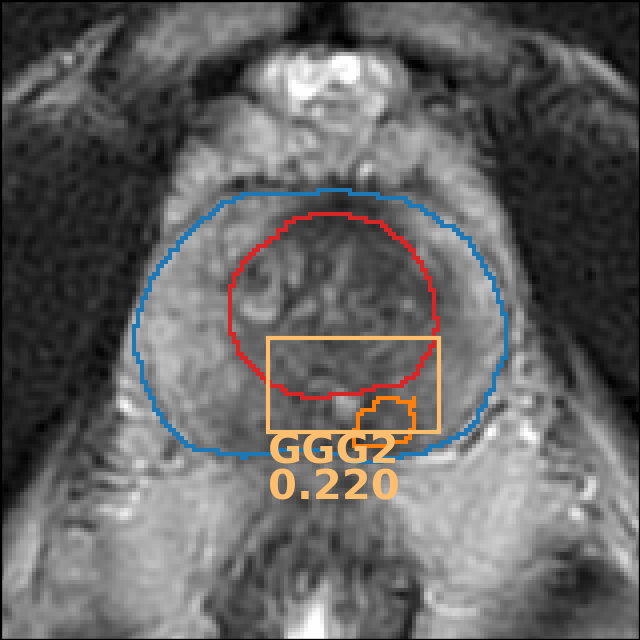}\hfill
		\includegraphics[width=0.198\linewidth]{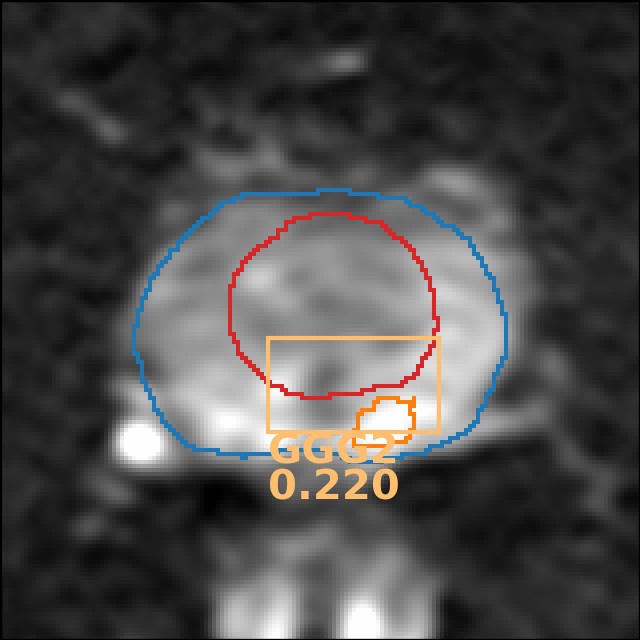}\hfill
		\includegraphics[width=0.198\linewidth]{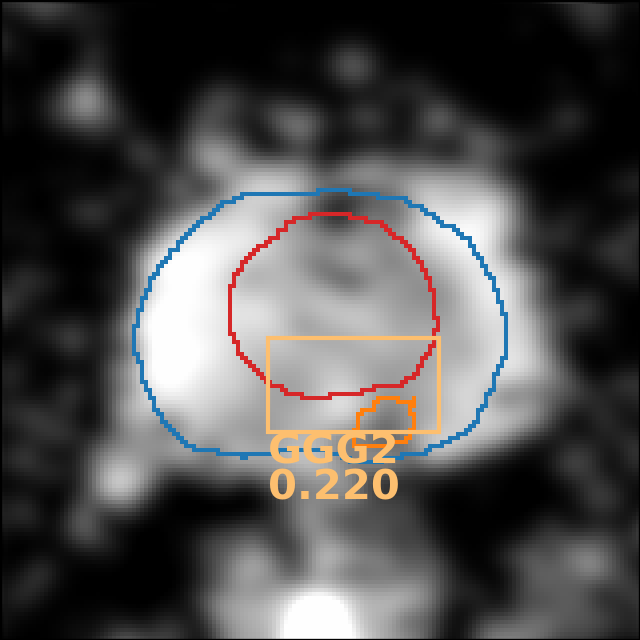}\hfill
		\includegraphics[width=0.198\linewidth]{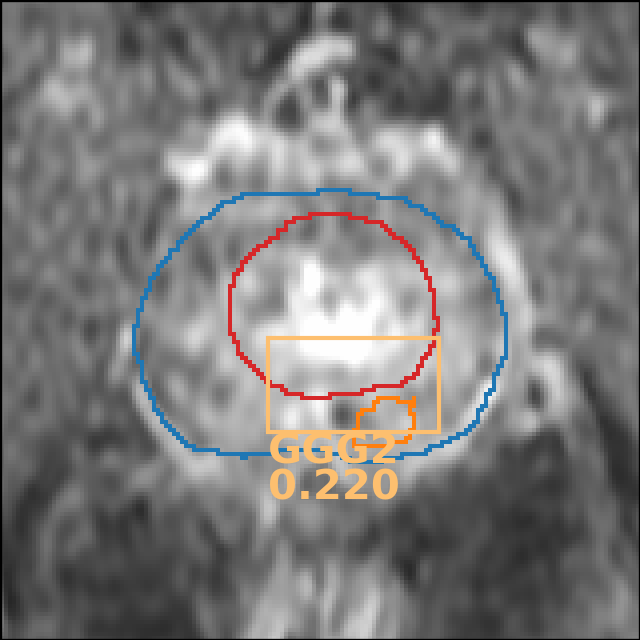}\\
		\vspace{0.5mm}
		\includegraphics[width=0.198\linewidth]{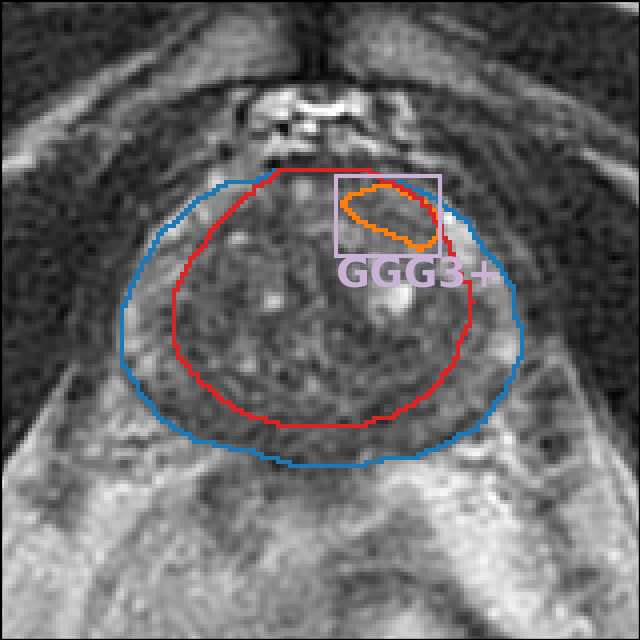}\hfill
		\includegraphics[width=0.198\linewidth]{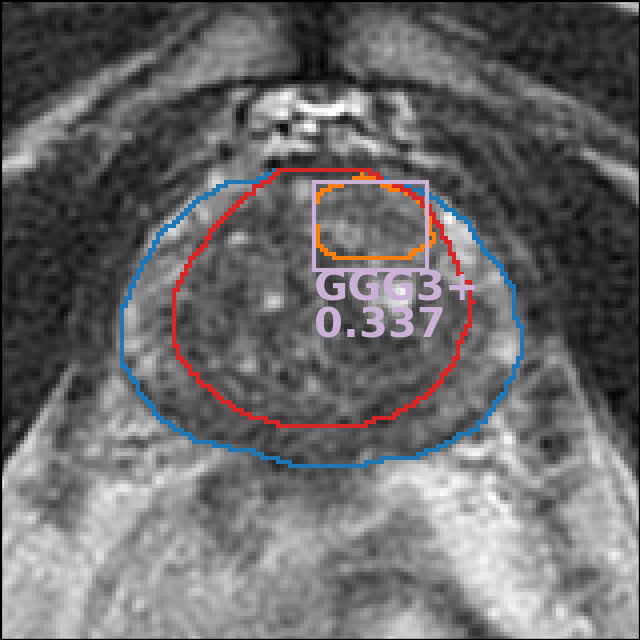}\hfill
		\includegraphics[width=0.198\linewidth]{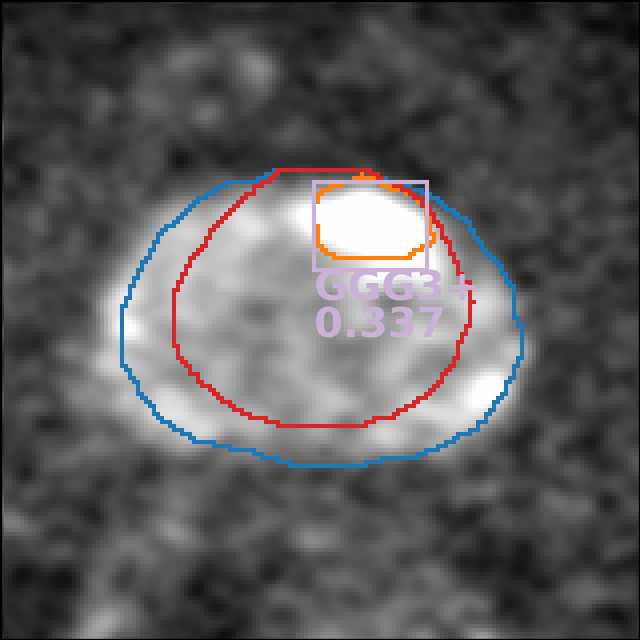}\hfill
		\includegraphics[width=0.198\linewidth]{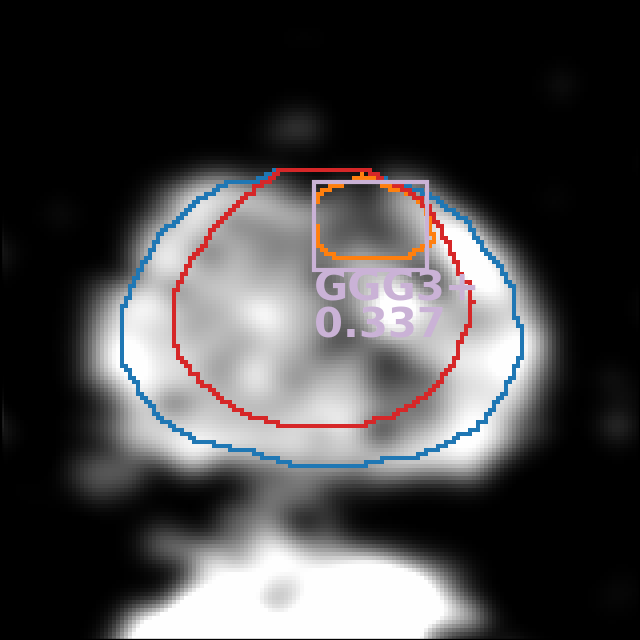}\hfill
		\includegraphics[width=0.198\linewidth]{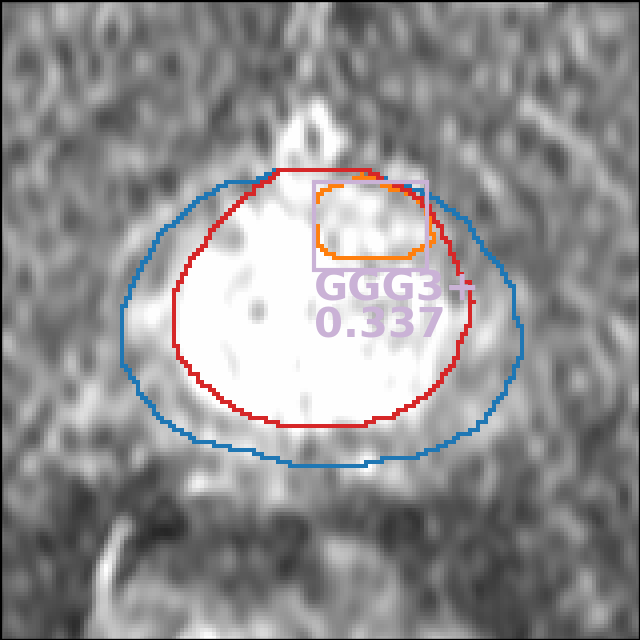}\\
		
	\end{widepage}
	\caption{Output of the model evaluated on two IVO test patients. The first image from the left shows the GT on the T2 sequence; the rest show the output predictions of the model on top different sequences (from left to right: T2, b1000/b1400, ADC, DCE $t=30$). GGG0 (benign) BBs are not shown and only the highest-scoring BB is shown for highly overlapped detections (IoU $>0.25$). Detections with a confidence below the lesion-wise maximum sensitivity setting ($t = 0.173$) are also ignored.}\label{fig:model_output}
\end{figure*}

\begin{figure*}[tb]
	\centering
	\begin{widepage}
		\includegraphics[width=0.198\linewidth]{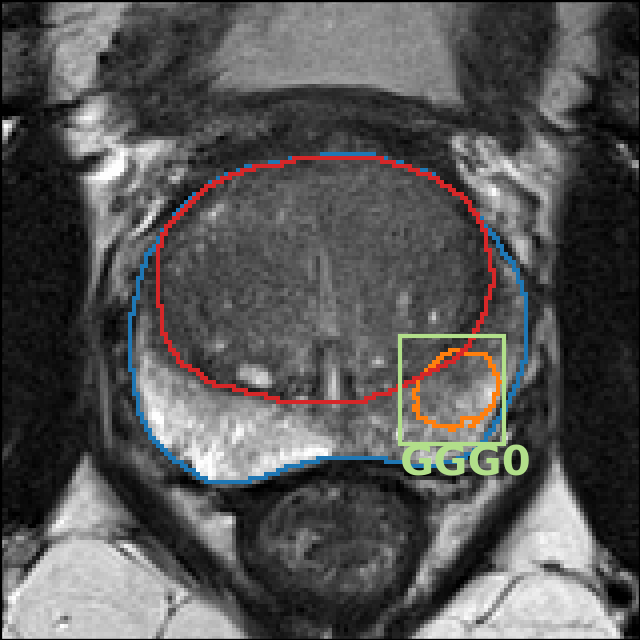}\hfill
		\includegraphics[width=0.198\linewidth]{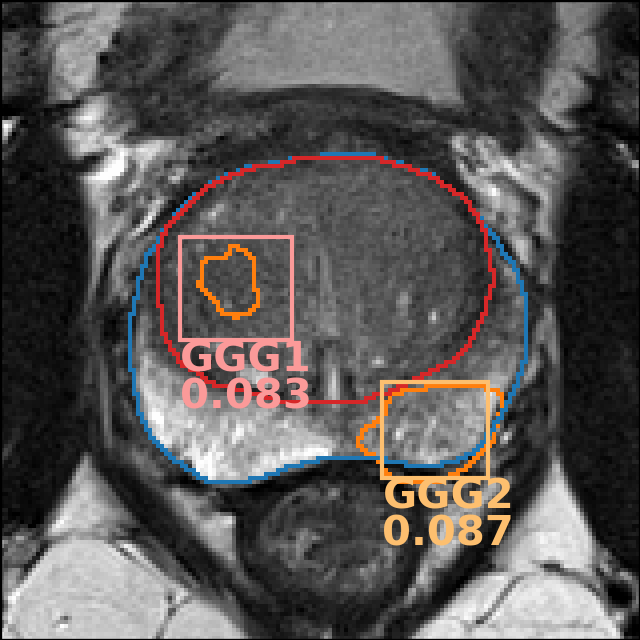}\hfill
		\includegraphics[width=0.198\linewidth]{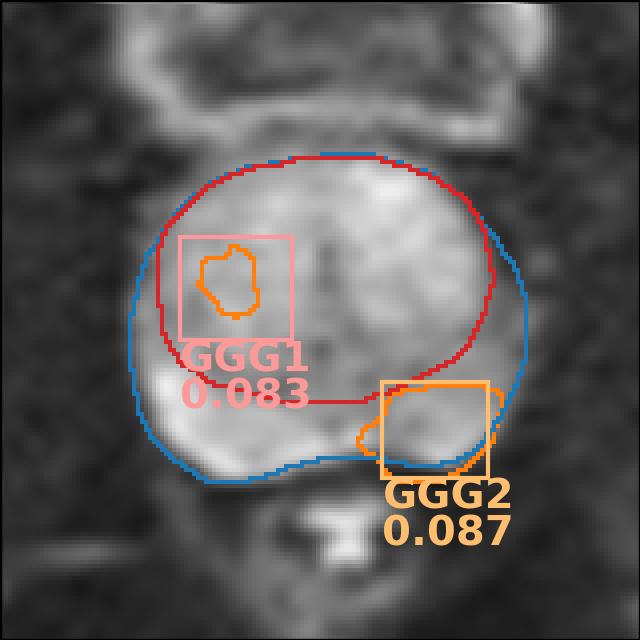}\hfill
		\includegraphics[width=0.198\linewidth]{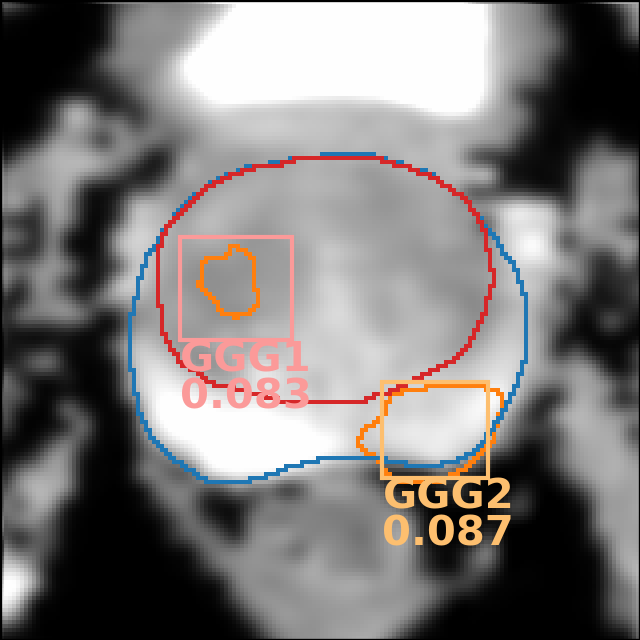}\hfill
		\includegraphics[width=0.198\linewidth]{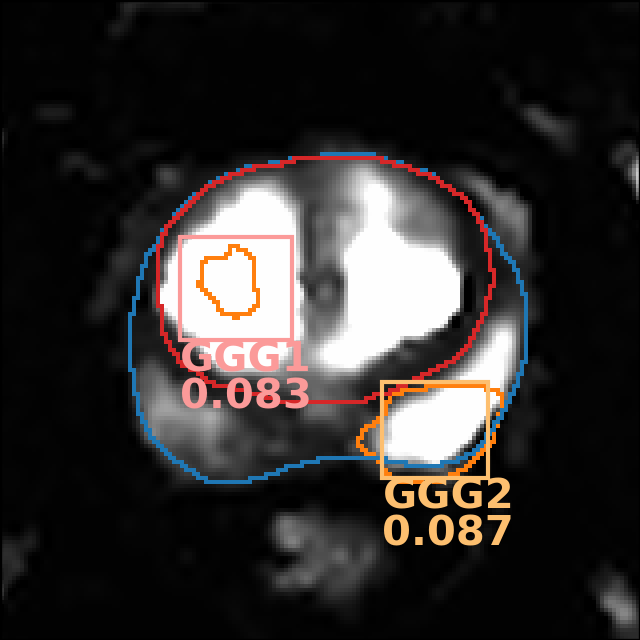}\\
		\vspace{0.5mm}
		\includegraphics[width=0.198\linewidth]{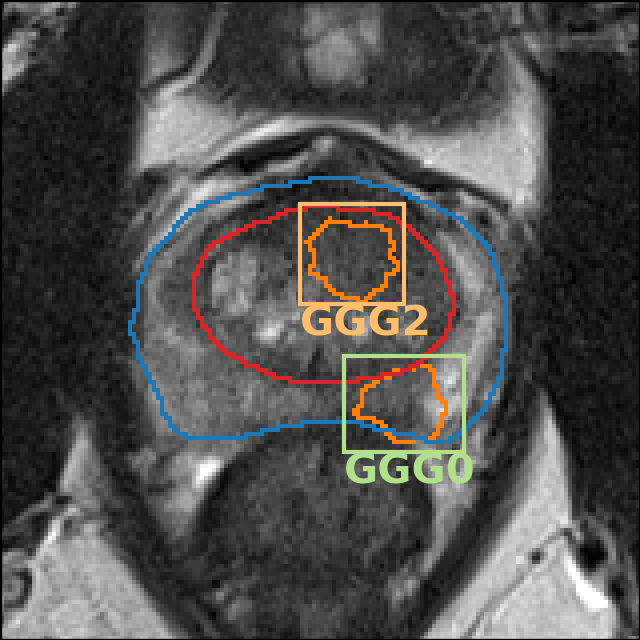}\hfill
		\includegraphics[width=0.198\linewidth]{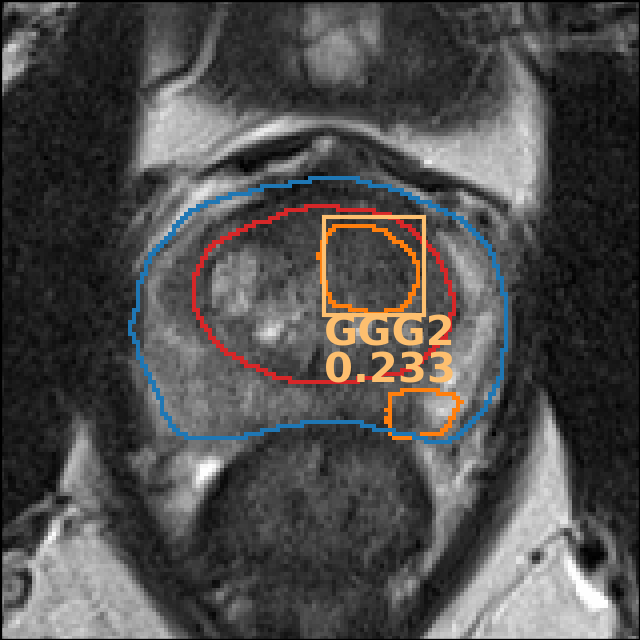}\hfill
		\includegraphics[width=0.198\linewidth]{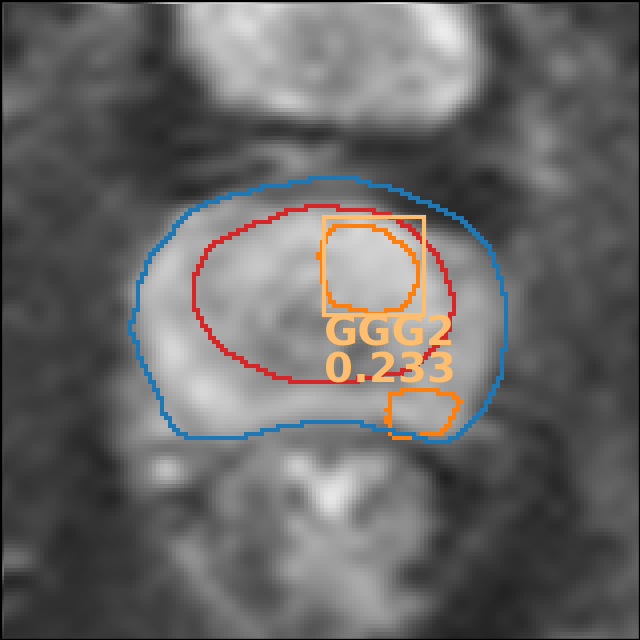}\hfill
		\includegraphics[width=0.198\linewidth]{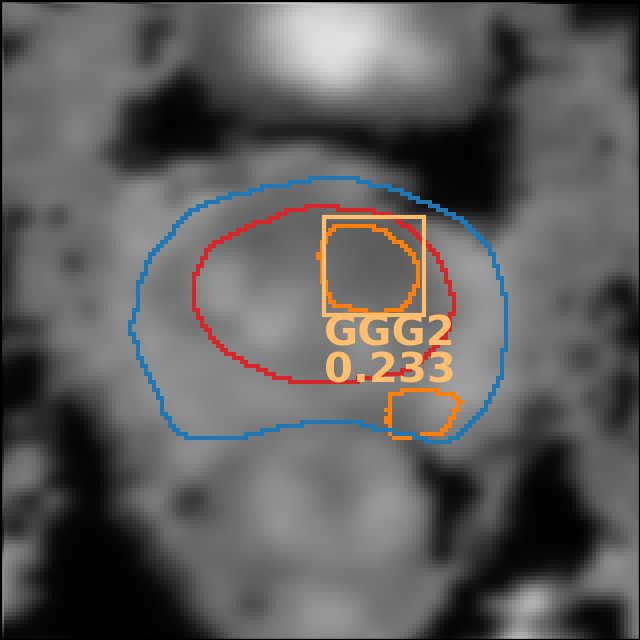}\hfill
		\includegraphics[width=0.198\linewidth]{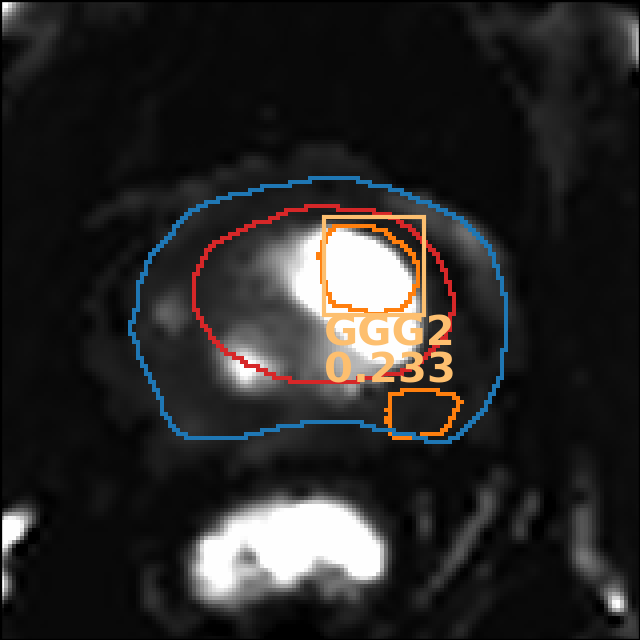}\\
		\vspace{0.5mm}
		\includegraphics[width=0.198\linewidth]{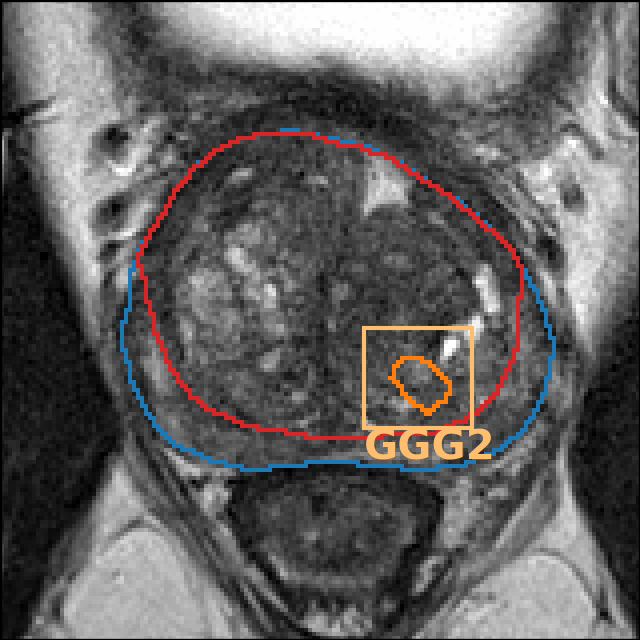}\hfill
		\includegraphics[width=0.198\linewidth]{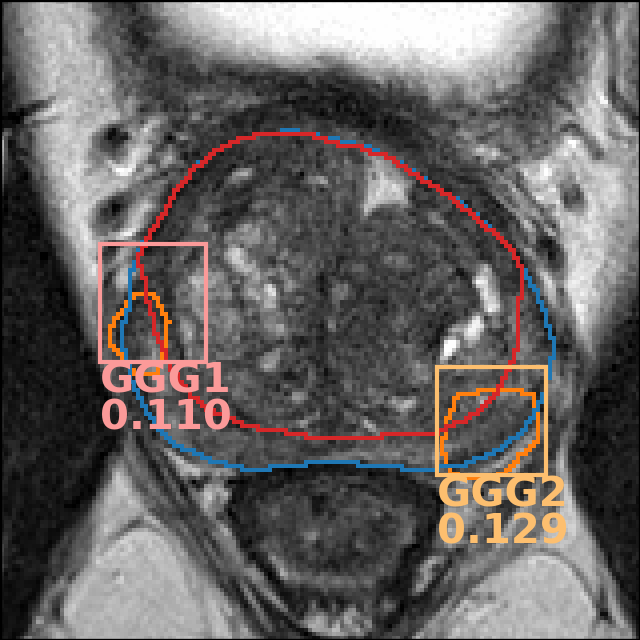}\hfill
		\includegraphics[width=0.198\linewidth]{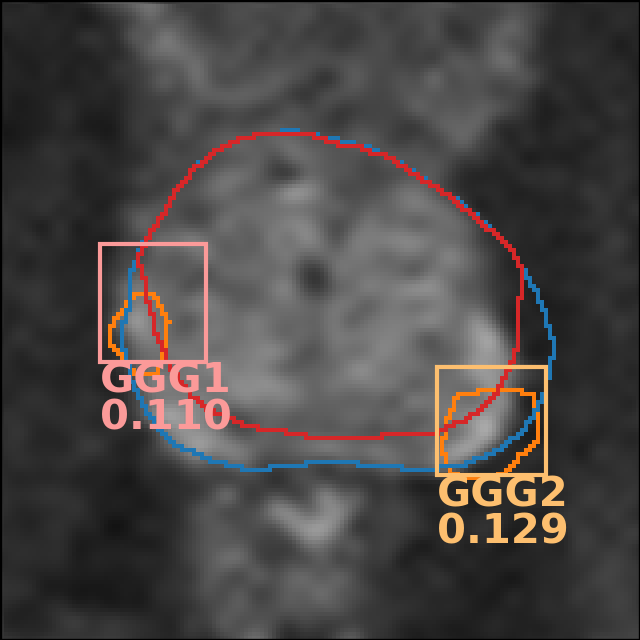}\hfill
		\includegraphics[width=0.198\linewidth]{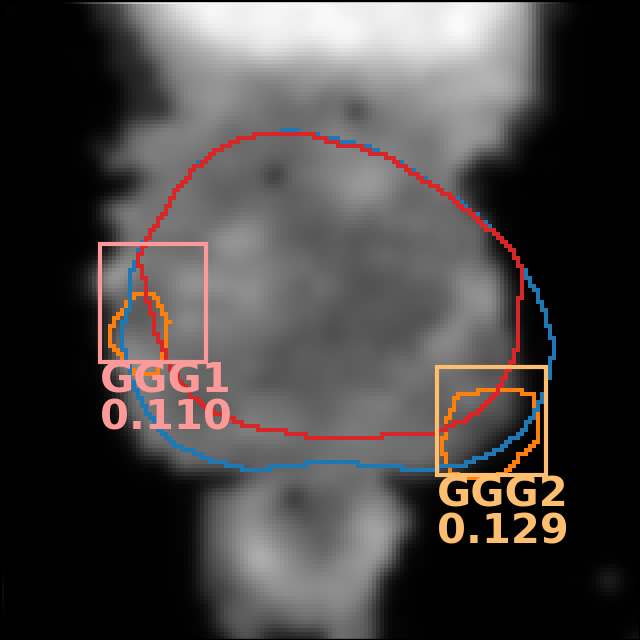}\hfill
		\includegraphics[width=0.198\linewidth]{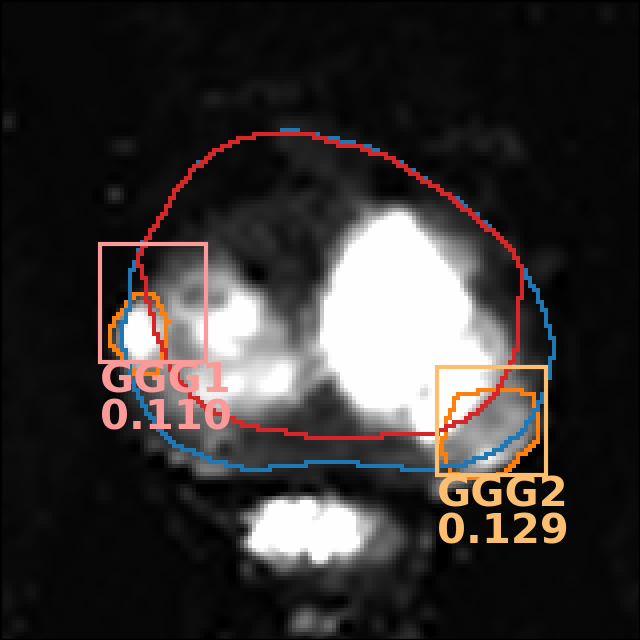}\\
		
	\end{widepage}
	\caption{Output of the model evaluated on three ProstateX test patients. First image from the left shows the GT on the T2; the rest show the output predictions of the model on different sequences (from left to right: T2, b800, ADC, $K^{trans}$.}\label{fig:model_output2}
\end{figure*}

Finally, Figure~\ref{fig:model_output} shows the output of the model evaluated on two IVO test patients and, Figure~\ref{fig:model_output2}, on three ProstateX test patients. Both figures (and several others in this paper) were generated using Python library plot\_lib\cite{Pellicer-Valero2020a}. For the sake of clarity, GGG0 (benign) BBs are not shown and, for highly overlapped detections (IoU \(> \ \)0.25), only the highest-scoring BB is drawn. Detections with confidence below the lesion-wise maximum sensitivity setting are also ignored. The first IVO patient (Figure~\ref{fig:model_output} top) is of special interest, as it is one of the relatively few IVO cases where the targeted biopsy did not find csPCa (as evidenced by the GGG0 BB in the GT image to the left), but the massive biopsy (20-30 cylinders) detected GGG2 csPCa. As can be seen, the model was able to detect this GGG2 lesion while ignoring the benign GGG0 one, hence outperforming the radiologists for this particular instance. For the second IVO patient (Figure\ref{fig:model_output} bottom) a GGG3+ GT lesion (GGG4 specifically) was properly detected by the model with very high confidence.

The first ProstateX patient (Figure~\ref{fig:model_output2} top) is a case of failure, where the model detects two non-existent GGG1 and GGG2 lesions, albeit with relatively low confidence; in fact, both would have been ignored at the balanced sensitivity setting ($t = 0.108$). For the next patient (Figure~\ref{fig:model_output2} middle), the model has been able to segment both GT lesions; however, only the csPCa lesion is detected, while the other is ignored (actually, the model correctly detected the other lesion as a GGG0, but BBs for those lesions are not shown). For the third patient (Figure~\ref{fig:model_output2} bottom), the model could correctly identify the GGG2 GT lesion but also identified an additional GGG1 lesion. This might be a mistake or might show a real lesion that was missed by the radiologists (we cannot know, as no massive biopsy information is available for the ProstateX dataset). Due to this uncertainty, lesion-level evaluation should not penalize detections for which GT information was not available (such as this one), as discussed in Section~\ref{lesion-matching-and-evaluation}.

\subsubsection{Sequence ablation tests}\label{sequence-ablation-tests}

In Section~\ref{online-data-augmentation}, Random Channel Drop is presented as a training-time data augmentation technique that should help alleviate the problem of missing sequences. For a model trained in such a fashion, we can assess the individual importance of the different sequences by dropping them (i.e.: setting them to 0) at test time and analyzing the performance penalty that the model incurs. The AUCs after dropping different sequences (or combinations of them) are shown in Table~\ref{tab:results_ablation}.

\begin{table}[h]
	\renewcommand{\arraystretch}{1.1}
	\caption{AUC after dropping one (or several) particular sequences (i.e.: setting it to 0) in test time for the GGG$\geq$2 significance criterion.}\label{tab:results_ablation}
	\small
	\begin{widepage}
	\centering
	\begin{tabular}{lllll}
		\toprule
		\thead{\multirow{2.5}{*}{\makecell{Sequence\\dropped}}} 
		& \multicolumn{2}{c}{\thead{ProstateX}} & \multicolumn{2}{c}{\thead{IVO}}\\
		\cmidrule(lr){2-3}
		\cmidrule(lr){4-5}
		& Lesion & Patient & Lesion & Patient\\
		\midrule
		None (Baseline) & 0.959 & 0.865 & 0.945 & 0.910\\
		b400/500 & 0.944 & 0.861 & 0.940 & 0.868\\
		b800/1000/1400 & 0.946 & 0.873 & 0.895 & 0.783\\
		All b-numbers & 0.951 & 0.844 & 0.845 & 0.720\\
		ADC & 0.905 & 0.870 & 0.940 & 0.836\\
		$K^{trans}$ & 0.894 & 0.865 & - & -\\
		All DCE & - & - & 0.895 & 0.820\\
		All but T2 & 0.804 & 0.808 & 0.782 & 0.545\\
		\bottomrule
	\end{tabular}
	\end{widepage}
\end{table}

As can be seen, removing the low b-valued (b400 for ProstateX/b500 for IVO) DW sequence seems to have minimal impact on both datasets, as is to be expected. Conversely, while removing the high b-valued (b800 for ProstateX/b1000 or b1400 for IVO) DW sequences has little impact on the ProstateX data, it severely affects the performance on the IVO data, likely due to the higher b values employed in this dataset (which may prove more informative). Furthermore, removing all DW sequences severely affects the IVO dataset, but has almost no impact on ProstateX. The removal of the ADC map has a similar negative impact on both datasets, although the results vary depending on how they are analyzed (lesion- or patient-wise). Likewise, dropping the $K^{trans}$ sequence on the ProstateX data or the DCE sequences on the IVO data clearly harms the performance. For the final test, all sequences are dropped except for the T2; despite it, the model still has a commendable performance, especially in the ProstateX set, which might indicate that the proposed Random Channel Drop augmentation has served its purpose of making the model more robust to missing sequences.

\section{Discussion}\label{sec:discussion}

Despite mpMRI interpretation being time-consuming and observer-dependent, it is a major clinical decision driver and poses great clinical relevance. In this paper we presented a CAD system developed with two main MRI datasets integrating T2, DW, b-value, and ADC maps in both of them as well as $K^{trans}$ for ProstateX and DCE for the IVO dataset. These were compared against fusion and transperineal template biopsies, which is considered the pre-operative gold standard to evaluate prostate cancer extent\cite{Drost2019}.

Different outcomes can be measured for this system. Firstly, regarding prostate zonal segmentation, we observed a great concordance between the model's and expert radiologist's prostate segmentation with a DSC that ranged from 0.894 to 0.941 depending on the MRI dataset. As can be seen, the results in the Private test set are extremely good, better in fact than any other model in the literature when evaluated in its internal test set and when evaluated blindly in the NCI-ISBI dataset. In Qin et al.\cite{Qin2020}, for instance, the authors train one CNN on an internal dataset and another identical CNN on the NCI-ISBI train dataset independently, and evaluate them by cross-validation, achieving a DSC of 0.908 and 0.785 at the CG and PZ in their internal dataset, and a DSC of 0.901 and 0.806 in the NCI-ISBI dataset. For a fairer comparison with our model, in Rundo et al.\cite{Rundo2019}, the authors train their model on two internal datasets (achieving a DSC of 0.829/0.859 in CG segmentation, and 0.906/0.829 in PZ segmentation), which then test blindly in the NCI-ISBI dataset, achieving 0.811 and 0.551 in CG and PZ segmentation, respectively. Finally, in Aldoj et al.\cite{Aldoj2020a}, training on a larger cohort of 141 patients and evaluating in their internal test set of 47, they achieve a DSC of 0.921, 0.895, and 0.781 for whole gland, CG, and PZ segmentation.

Regarding lesion detection as exposed in Section~\ref{quantitative-results}, the results for lesions GGG$\geq$2 significance criterion can be considered as optimal: all csPCa lesions were detected while maintaining a very high specificity, except for the patient-level ProstateX evaluation, and a great AUC ranging from 0.865 to 0.959. Furthermore, the IVO results outperform the PI-RADS scores, especially at the high sensitivity setting (PI-RADS$\geq 4$) which is of most interest in clinical practice. This can be seen in Figure~\ref{fig:auc}, where the ROC is at all instances above and to the left of the PI-RADS scores. For further comparison, several studies have reported radiologist sensitivities/specificities for the detection of csPCa from mpMRI at a patient level of 0.93/0.41\cite{Ahmed2017}, or 0.58-0.96/0.23-87 as shown in a systematic review\cite{Futterer2015}. The results vary wildly due to their single-center nature, their differing criteria for the definition of csPCa, and the often-inaccurate reference standards employed.

Considering GGG$\geq$3 significance criterion, caution is required when interpreting these results due to the very low number of positive cases (e.g.: only 3 in the IVO test set). Furthermore, the 0.714 patient-level sensitivity does not mean that the model missed GGG3 lesions, but rather that they were assigned to a lower GGG (such as GGG2) and were therefore ignored for the GGG$\geq$3 classification problem.

In addition to the previous tests, the ongoing ProstateX challenge was used for external lesion-level validation, achieving an AUC of 0.85, which would have been the second-best AUC in the original ProstateX challenge\cite{Armato2018}. Additionally, an identical model trained only on the ProstateX data (which has been made publicly available alongside this paper), achieved an AUC of 0.87, which would have tied with the best contender in the challenge. There are now higher AUCs in the online leaderboard but, unfortunately, we were unable to find any publications regarding them, and hence no further analysis can be performed. In any case, these results must be also interpreted with caution: on one hand, the proposed system solves a much more complex problem (namely detection, segmentation \& classification) than the comparatively simpler ROI classification systems which are typically employed for this task, and it is therefore in a disadvantage compared to them. On the other hand, as indicated in Section~\ref{data-description}, the ProstateX challenge mpMRIs were used for training the segmentation and detection components of the model, but not the classification head (as GGG information is kept secret by the challenge, and hence unavailable for training). The inclusion of this data was useful for increasing the number of training samples, although it might have introduced some unknown bias for the evaluation of this dataset.

Outside the ProstateX challenge, one of the very first works on the topic by Litjens et al.\cite{Litjens2017} reported a sensitivity of 0.42, 0.75, and 0.89 at 0.1, 1, and 10 false positives per normal case using a classical radiomics-based model. More recently, Xu et al.\cite{Xu2019} used a csPCa segmentation CNN whose output was later matched to GT lesions based on distance (similar to ours). He reported a sensitivity of 0.826 at some unknown specificity; also, despite using the ProstateX data, unfortunately, no ProstateX challenge results were provided. Cao et al.\cite{Cao2019} proposed a segmentation CNN that also included GGG classification as part of its output, reporting a maximum sensitivity of 0.893 at 4.64 false positives per patient and an AUC of 0.79 for GGG$\geq$2 prediction. Interestingly, the authors employed histopathology examinations of whole-mount specimens as GT for the model. Aldoj et al.\cite{Aldoj2020} utilized the ProstateX data to perform csPCa classification on the mpMRI ROIs around the provided lesion positions, reporting an AUC of 0.91 on their internal 25-patient test set; once again, despite using the ProstateX data exactly as conceived for the challenge, they do not provide any challenge results for comparison.

In an interesting prospective validation study, Schelb et al.\cite{Schelb2020} obtained a sensitivity/specificity of 0.99/0.24 using a segmentation CNN, a performance that they found comparable to radiologist-derived PI-RADS scores. Woźnicki et al.\cite{Woznicki2020} proposed a classical radiomics-based model (no CNNs involved) achieving an AUC of 0.807. Finally, as for patient-level csPCa classification results, Yoo et al.\cite{Yoo2019} achieved an AUC of 0.84 using slice-wise CNN classifier whose predictions were later combined into a patient-wise total score and Winkel et al.\cite{Winkel2020} achieved a sensitivity/specificity of 0.87/0.50 on a prospective validation study using a segmentation-based detection system which is most similar to the one proposed here.

Considering lesion segmentation concordance, as exposed in Section~\ref{quantitative-results}, our results are unfortunately not directly comparable to other papers in the literature, as those focus on segmenting exclusively csPCa and benign lesions are ignored. For instance, Schelb et al.\cite{Schelb2020} reported a DSC of 0.34 for csPCa segmentation, similar to Vente et al.\cite{Vente2021}'s 0.37 DSC\textbf{.} When interpreting segmentation results it is important to bear in mind that ProstateX segmentations are being compared against an automatically generated segmentation, which are far from being a perfect GT. Secondly, it is challenging and of dubious relevance to compare lesion segmentation concordance as mpMRI lesions tend to be small with ill-defined margins and the inter-observer variability is high\cite{Steenbergen2015}. Lastly, the main output of the model is the detection and categorization of the lesions rather than their accurate segmentation.

AI mpMRI interpretation represents a very promising line of research that has already been successfully applied to prostate gland segmentation and PCa lesion detection using both transperineal prostate biopsy and radical prostatectomy specimens as GT with varying results\cite{Yoo2019,Winkel2020}. We went a step further and developed the first algorithm, to the best of our knowledge, that automatically contours the prostate into its zones, performs well at lesion detection and Gleason Grade prediction (identifying lesions of a given grade or higher), and segments such lesions albeit with a moderate overlapping. The model outperformed expert radiologists with extensive MRI experience and achieved top results in the ProstateX challenge.

The code has been made available, including an automatic prostate mpMRI non-rigid registration algorithm and an automatic mpMRI lesion segmentation model. Most importantly, the fact that the code is online might allow future researchers to use this model as a reference upon which to build or to compare their models.

Our work presents some limitations. Firstly it would require further validation and prospective blinded trial to assess histological results of targeted biopsies to the lesions identified by the model. Secondly, although the model was successfully trained on two datasets, it still behaves differently on each of them (e.g.: the optimal thresholds vary significantly between them), which is not desirable, but probably unavoidable.

In any case, this is yet another step in the foreseeable direction of developing a strong collaborative AI net that progressively incorporates as many mpMRIs with the corresponding GT as possible. The clinical applications of this model are countless, amongst which we could consider assisting radiologists by speeding up prostate segmentation, training purposes as well as a safety net to avoid missing PCa lesions. Further, the ability to detect csPCa can easily highlight which MRIs would require prompt reporting and prioritizing biopsy. Moreover, given the recent trend towards conservative PCa approaches such as focal therapy or active surveillance (usually implying a more dedicated prostate biopsy), predicting the Gleason Grade, as well as the number of lesions pre-biopsy, could identify eligible men that could be offered transperineal targeted biopsy in the first place.

\section*{Author contributions}\label{author-contributions}

Conceptualization, José D. Martín-Guerrero; Data curation, Oscar J. Pellicer-Valero and Victor Gonzalez-Perez; Methodology, Oscar J. Pellicer-Valero; Project administration, José D. Martín-Guerrero; Resources, Victor Gonzalez-Perez, Isabel García and María Benito; Software, Oscar J. Pellicer-Valero; Supervision, Juan Luis Ramón-Borja, José Rubio-Briones, María Rupérez and José D. Martín-Guerrero; Validation, Isabel García, María Benito and Paula Gómez; Visualization, Oscar J. Pellicer-Valero; Writing -- original draft, Oscar J. Pellicer-Valero, José L. Marenco Jiménez; Writing -- review \& editing, Victor Gonzalez-Perez, María Rupérez and José D. Martín-Guerrero.

\section*{Funding}\label{funding}

This work has been partially supported by a doctoral grant of the Spanish Ministry of Innovation and Science, with reference FPU17/01993.

\section*{Institutional review}\label{institutional-review}

The study was approved by the Ethical Committee of the Valencia Institute of Oncology (CEIm-FIVO) with protocol code PROSTATEDL (2019-12) and date 17\textsuperscript{th} of July, 2019.

\section*{Data availability}\label{data-availability}

Data from the ProstateX challenge are available at \url{https://doi.org/10.7937/K9TCIA.2017.MURS5CL}\cite{Litjens2017}; data from the Valencian Institute of Oncology is not publicly available, since the ethical committee (CEIm-FIVO) only approved its use for the current study. They might be made available for research purposes on reasonable request from the corresponding author. The code of the project is available at \url{https://github.com/OscarPellicer/prostate_lesion_detection}.

\section*{Conflicts of interest}
The authors declare no conflict of interest.

\onecolumn

\end{document}